\documentclass[prb,showpacs,superscriptaddress,onecolumn]{revtex4-2}
\usepackage[utf8]{inputenc}
\usepackage{amsmath,graphicx,hyperref}
\usepackage{xcolor}

\allowdisplaybreaks[1]
\setcitestyle{super}

\newcommand*{\citen}[1]{%
  \begingroup
    \romannumeral-`\x 
    \setcitestyle{numbers}%
    \cite{#1}%
  \endgroup   
}

\usepackage{setspace}

\begin{document}

\doublespacing

\makeatletter
\let\ps@titlepage\ps@plain
\makeatother

\title{Strain as a tool to stabilize the isotropic triangular lattice in a geometrically frustrated organic quantum magnet}

\author{Francisco Lieberich}
\affiliation{Max Planck Institute for Chemical Physics of Solids, 01187 Dresden, Germany}
\affiliation{Technical University of Dresden, 01187 Dresden, Germany}

\author{Yohei Saito}
\affiliation{Institute of Physics, Goethe University Frankfurt, 60438 Frankfurt (M), Germany}

\author{Yassine Agarmani}
\affiliation{Institute of Physics, Goethe University Frankfurt, 60438 Frankfurt (M), Germany}

\author{Takahiko Sasaki}
\affiliation{Institute for Materials Research, Tohoku University, Sendai 980-8577, Japan}

\author{Naoki Yoneyama}
\affiliation{Graduate Faculty of Interdisciplinary Research, University of Yamanashi, Kofu 400-8511, Japan}

\author{Stephen M. Winter}
\affiliation{Department of Physics and Center for Functional Materials, Wake Forest University, Winston-Salem, North Carolina, 27109, USA}

\author{Michael Lang}
\affiliation{Institute of Physics, Goethe University Frankfurt, 60438 Frankfurt (M), Germany}

\author{Elena Gati}
\email{elena.gati@cpfs.mpg.de}
\affiliation{Max Planck Institute for Chemical Physics of Solids, 01187 Dresden, Germany}
\affiliation{Technical University of Dresden, 01187 Dresden, Germany}

\date{\today}

\begin{abstract}
Geometric frustration is a key ingredient in the emergence of exotic states of matter, such as the quantum spin liquid in Mott insulators. While there has been intense interest in experimentally tuning frustration in candidate materials, achieving precise and continuous control has remained a major hurdle — particularly in accessing the properties of the ideally frustrated lattice. Here, we show that large, finely controlled anisotropic strains can effectively tune the degree of geometric frustration in the Mott insulating $\kappa$-(ET)$_2$Cu$_2$(CN)$_3$ -- a slightly anisotropic triangular-lattice quantum magnet. Using thermodynamic measurements of the elastocaloric effect, we experimentally map out a temperature-strain phase diagram that captures both the ground state of the isotropic lattice and the less frustrated parent state. Our results provide a new benchmark for calculations of the triangular-lattice Hubbard model as a function of frustration and highlight the power of lattice engineering as a route to realizing perfectly frustrated quantum materials.
\end{abstract}



\maketitle

\newpage

\subsection*{Introduction}

\noindent Frustration in an interacting quantum system refers to the incompatibility of interactions that cannot all be satisfied simultaneously. It is widely accepted that frustration is the source of a number of exotic phenomena in quantum materials. Most prominently, in magnetic insulators frustration suppresses the tendency of interacting spins to establish long-range order. Instead of undergoing a symmetry-breaking phase transition, strong quantum fluctuations cause the electron spins to remain disordered in a liquid-like fashion even at absolute zero temperature. Such quantum spin liquids (QSL)~\cite{Balents10,Zhou17,Broholm20} exhibit highly unusual and exciting properties, such as long-range quantum entanglement and fractionalized excitations. A common approach to induce such exotic states is through geometric frustration~\cite{Moessner06}, where the atomic arrangement itself prevents the minimization of all interaction energies. This effect occurs, for instance, on a triangular lattice, where three neighboring spins cannot all align antiferromagnetically. Moreover, geometric frustration not only induces non-trivial behavior of spins, but also may give rise to new states of matter with intriguing static and dynamic properties resulting from charge degrees of freedom~\cite{Kagawa13,Sasaki20}. For example, in metals, hopping frustration on triangular-lattice variants can lead to 'flat bands'~\cite{Ye24,Wilson24}, which may host novel correlated states and topological phases.

The wide range of phenomena predicted to arise from frustration has led to intense experimental efforts to realize and identify these effects. However, achieving precise and clean control of geometric frustration in real materials remains a significant challenge~\cite{Lacroix11,Broholm20}. To date, most tuning approaches rely on chemical modification, which commonly introduces disorder into the system. Such disorder can obscure the intrinsic properties of interest and complicates the interpretation of results. In magnetic insulators, for instance, disorder can lead to a magnetic state without long-range order that is not a true spin liquid. As a result, methods that allow precise tuning of frustration in a single, pristine sample — without introducing additional disorder — are highly desirable. Such approaches would enable fundamental open questions in the field of frustrated magnetism to be addressed, including: Is the ground state of a spin-liquid candidate, which is typically strongly but not perfectly frustrated, also the ground state at optimal frustration, or do other phases emerge? Does a candidate spin-liquid system develop magnetic order upon controlled reduction of frustration or does it remain disordered? If a transition occurs, what is the nature of the parent phase and of the phase transition\cite{Senthil04}? Conversely, can frustration in a moderately frustrated system be enhanced to induce spin-liquid behavior?

\begin{figure} 
	\centering
	\includegraphics[width=\textwidth]{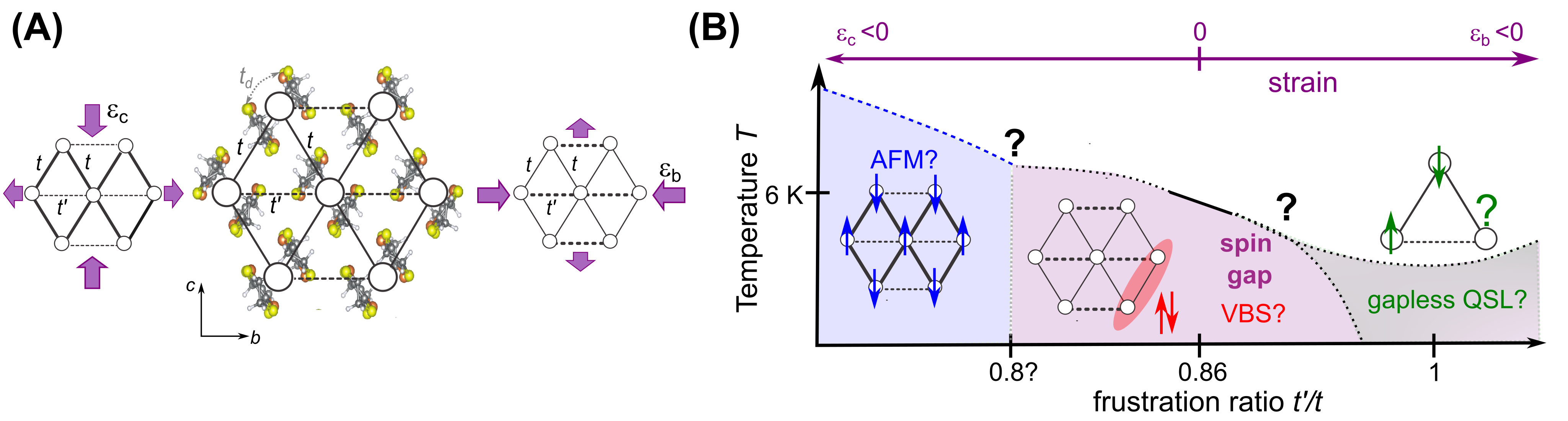} 
	\caption{\textbf{Strain tuning of a geometrically-frustrated triangular lattice with hopping parameters $t$ and $t^\prime$.} (\textbf{A}) Uniaxial pressure induces anisotropic strains (indicated by purple arrows), which vary depending on the sign and direction of the applied pressure. These strains, denoted as $\varepsilon_\textrm{c}$ and $\varepsilon_\textrm{b}$, change the degree of frustration, quantified by the ratio $t^\prime/t$.  Compression along the $c$-axis ($b$-axis) is expected to decrease (increase) $t^\prime/t$. The material $\kappa$-(ET)$_2$Cu$_2$(CN)$_3$ (ET=bisethylenedithio-tetrathiafulvalene), whose crystal structure is depicted in the middle, is an excellent material realization of an almost isotropic triangular lattice. In this material, (ET)$_2^+$-dimers (represented by the white circles), each carrying a spin-1/2, are arranged on a frustrated triangular lattice with $t^\prime/t~=~0.86$~\cite{Kandpal09,Jeschke12}. The effective-dimer triangular-lattice model provides a good description when the intra-dimer hopping $t_d$ is large ($t_d~\gg~t, t^\prime$). \textbf{(B)} Expected phase diagram for the frustrated magnet as a function of frustration, $t^\prime/t$, which can only be accessed experimentally through continuous strain tuning. For $\kappa$-(ET)$_2$Cu$_2$(CN)$_3$ a spin gap opens below $T^\star~=6~$K~\cite{Miksch21,Manna10}, likely due to the formation of a valence-bond solid (VBS) state. Whereas the change of $T^\star$ with small changes of $t^\prime/t$ can be inferred from thermodynamic measurements at ambient pressure~\cite{Manna18} (black solid line), the behavior beyond the threshold of perturbatively small changes of $t^\prime/t$ is the focus of the present study (black dashed lines).  Key questions relate to the properties of $\kappa$-(ET)$_2$Cu$_2$(CN)$_3$ on the isotropic lattice, $t^\prime/t~=~1$: Do possible gapless quantum spin-liquid (QSL) states emerge? What is the nature of the parent states at lower frustration, such as antiferromagnetic (AFM) phases or other ordered states? }
	\label{fig:overview} 
\end{figure}

In this work, we show that experimentally-achievable, large  anisotropic strains allow clean control of the frustration in a real candidate material (see Fig.~\ref{fig:overview}~\textbf{A}). We take advantage of technical advances in order to tune quantum materials by strain using piezoactuator-driven uniaxial pressure cells~\cite{Barber19}. We combine this with elastocaloric effect (ECE) measurements, recently pioneered in the context of unconventional superconductors~\cite{Ikeda19,Li22}, and introduce ECE measurements as a powerful thermodynamic tool for high-resolution mapping of the phase diagram of frustrated systems under strain. Specifically, we report what is, to the best of our knowledge, the first experimentally-determined thermodynamic temperature-strain phase diagram of a frustrated triangular-lattice quantum magnet. By both increasing and decreasing frustration with strain, we capture the properties of the ground state at maximal frustration as well as a transition to a parent state at lower frustration. Our experimental results under continuous strain tuning open a new pathway for accessing and validating the rich phase space predicted by frustrated Hubbard models.

\subsection*{Results}

\noindent The frustrated Mott insulator $\kappa$-(ET)$_2$Cu$_2$(CN)$_3$ is well suited for this study, since it presents one of the rare realizations of a nearly isotropic triangular lattice~\cite{Powell11,Riedl22,Schaefer24,Kandpal09,Jeschke12} (see Fig.~\ref{fig:overview}~\textbf{A}). In this material, (ET)$_2$ dimers, each carrying spin-1/2, reside on a slightly anisotropic triangular lattice, as measured by the ratio of the hopping parameters  $t^\prime$ and $t$, with $t^\prime/t~\approx~0.86$~\cite{Kandpal09,Jeschke12}. Despite strong exchange interactions ($J/k_\textrm{B}~\approx~250$~K), the lack of long-range magnetic order down to 32~mK has positioned $\kappa$-(ET)$_2$Cu$_2$(CN)$_3$ as a promising QSL candidate~\cite{Shimizu03}, stabilized by ring-exchange interactions in weak Mott insulators. Recent experiments provided compelling evidence  that the anomalous behavior observed around $T^\star~\approx~6~$K in this compound (commonly referred to as the `6~K anomaly'~\cite{AbdelJawad10,Manna10,Poirier14,Yamashita08,Yamashita09,Isono16}) is related to the opening of a spin gap~\cite{Miksch21}. The latter finding has been interpreted in favor of a phase transition into a valence-bond solid (VBS) ground state~\cite{Riedl19,Miksch21,Matsuura22,Pustogow23}.

This material provides an excellent setting to explore the key questions, introduced above (see Fig.~\ref{fig:overview}~\textbf{B}). In order to address these questions for the case of a triangular-lattice quantum magnet, we increase and decrease frustration by applying uniaxial pressure to the fragile $\kappa$-(ET)$_2$Cu$_2$(CN)$_3$ crystals either along the crystallographic $b-$ or $c$-axis. The applied uniaxial pressure generates compressive strain along the pressure axis and smaller tensile strain along the perpendicular axes, determined by Poisson’s ratio. In this study, we denote these anisotropic strains as $\varepsilon_\textrm{i}$ ($i~=~b,c$), where $i$ refers to the direction of the applied pressure (see Fig.~\ref{fig:overview}~\textbf{A}). Whereas compressive $\varepsilon_\textrm{c}$ ($<~0$) brings $\kappa$-(ET)$_2$Cu$_2$(CN)$_3$ closer to the less frustrated square lattice geometry, compressive $\varepsilon_\textrm{b}$ ($<~0$) will tune the system closer to and possibly beyond the isotropic triangular limit (see Fig.~\ref{fig:overview}~\textbf{A}). The anisotropic strains are approximately volume-conserving, so that they have a much greater impact on the frustration strength than on the correlation strength~\cite{Shimizu11} (see Supplementary Information). 

The recently developed ECE technique promises to be a highly sensitive thermodynamic probe of frustrated magnetism under strain. The ECE, $\eta_\textrm{i}:=\left(\frac{\Delta T}{\Delta \varepsilon_\textrm{i}}\right)_S$, describes the adiabatic temperature change, $\Delta T$, induced by a change in strain, $\Delta \varepsilon_\textrm{i}$ (see Fig.~\ref{fig:raw-data}~\textbf{A}). The sensitivity of $\eta_\textrm{i}$ arises from its direct proportionality to strain-induced changes in entropy, which are expected to be large in a frustrated magnet. This connection becomes evident from the thermodynamic expression for $\eta_\textrm{i}$, which links it to the derivative of the entropy, $S$, with respect to strain, $\varepsilon$, and temperature, $T$:

\begin{equation}
    \eta_\textrm{i}= \left(\frac{\Delta T}{\Delta \varepsilon_\textrm{i}}\right)_S = -\frac{\left(\frac{\partial S}{\partial \varepsilon_\textrm{i}} \right)_T}{\left( \frac{\partial S}{\partial T}\right)_{\varepsilon_\textrm{i}}}.
    \label{eq:ECE}
\end{equation}

\noindent Furthermore, this equation shows that $\eta$ vanishes whenever the entropy is extremal~\cite{Li22,Gati23}. This is particularly important to directly identify experimentally the strains at which the entropy is maximal and frustration is optimal. This unique property distinguishes the ECE apart from other thermodynamic probes, such as specific heat, which is only sensitive to the temperature-dependence of the entropy.

\subsubsection*{The elastocaloric effect at zero and finite strains}

\begin{figure}
\centering
\includegraphics[width=\textwidth]{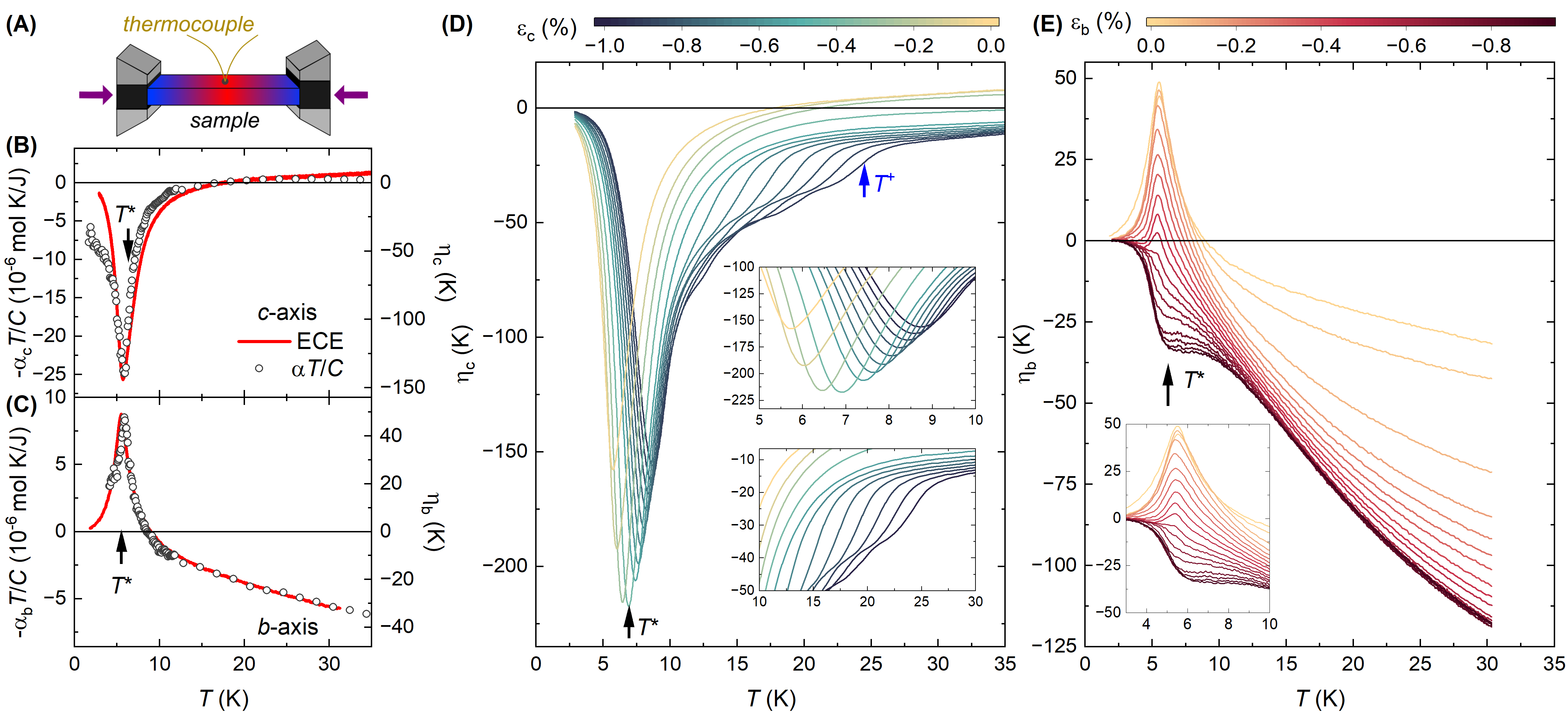} 
\caption{\textbf{Elastocaloric effect of $\kappa$-(ET)$_2$Cu$_2$(CN)$_3$ at zero and finite strains.} (\textbf{A}) The elastocaloric effect (ECE), $\eta$, refers to the temperature change, $\Delta T$, which is induced by a change of strain, $\Delta \varepsilon$, and is measured through a thermocouple attached to the middle of the sample~\cite{Ikeda19}. $\eta$ can be determined at different tuning strains $\varepsilon$. (\textbf{B},\textbf{C}) Thermodynamic considerations dictate that $\eta_\textrm{i}(\varepsilon_\textrm{i}~=~0)$, induced by $\Delta \varepsilon_\textrm{i}$ ($i~=~c,b$) is related to the reported ambient-pressure thermal expansion, $\alpha_\textrm{i}(T)$~\cite{Manna10}, and the specific heat, $C(T)$~\cite{Yamashita08} (see Supplementary Information). The `6~K-anomaly' at $T^\star$ is clearly resolved in the ECE data. (\textbf{D}) $\eta_\textrm{c}$ under finite, compressive $\varepsilon_\textrm{c} (<~0)$. At small compression, a clear shift of $T^\star$ is observed. For higher compression, a second anomaly at $T^+$ appears, that signals the transition into a new thermodynamic phase. (\textbf{E}) Elastocaloric $\eta_\textrm{b}$ under finite, compressive $\varepsilon_\textrm{b}$. Upon increasing compression, the anomalous contribution to $\eta$ around $T^\star$ reduces and eventually changes sign. Following thermodynamic considerations (see text), this implies that the strain dependence, d$T^\star$/d$\varepsilon_\textrm{b}$, changes sign at intermediate $\epsilon_b$, while $T^\star$ is not suppressed to zero temperature.}
\label{fig:raw-data}
\end{figure}

\noindent In contrast to the specific heat, $C$, or the thermal expansion, $\alpha_\textrm{i}$, the ECE, $\eta_\textrm{i}$, has not yet been studied for the family of organic-charge transfer salts, to which $\kappa$-(ET)$_2$Cu$_2$(CN)$_3$ belongs. To demonstrate that we probe the intrinsic $\eta_\textrm{i}$ of $\kappa$-(ET)$_2$Cu$_2$(CN)$_3$, we first compare our measured $\eta_\textrm{i}$ at near-zero strain with the $C$ and $\alpha_\textrm{i}$ data from literature~\cite{Yamashita08,Manna10,Manna18}, which clearly identified the `6~K anomaly'. Based on Eq.~\ref{eq:ECE}, $\eta_\textrm{i}$ is related to $-\alpha_\textrm{i} T/C$ (see Supplementary Information). In Figs.~\ref{fig:raw-data} \textbf{B} and \textbf{C}, we compare our $\eta_\textrm{i}$ that is measured when the a.c. strain is applied along the $i~=~b$-axis and the $c$-axis with $-\alpha_\textrm{b} T/C$ and $-\alpha_\textrm{c} T/C$, respectively. These plots show a strong agreement between the different thermodynamic quantities over the investigated temperature range. The extraordinarily sensitivity of the ECE in detecting the `6 K anomaly' at $T^\star$ is in accordance with the anticipated strong strain sensitivity of this frustrated magnet. Assuming that the anomaly at $T^\star$ corresponds to a true thermodynamic phase transition, the signatures in $\alpha_\textrm{i}$ and $\eta_\textrm{i}$ dictate that compressive $\varepsilon_\textrm{c}$ ($\varepsilon_\textrm{b}$) will initially increase (suppress) $T^\star$.

To determine the phase diagram under larger strain, we show in Fig. \ref{fig:raw-data}~ \textbf{D} and \textbf{E} our results of $\eta_\textrm{i}$ ($i=~b,c$) of $\kappa$-(ET)$_2$Cu$_2$(CN)$_3$ under finite compressive $\varepsilon_\textrm{c}$ and $\varepsilon_\textrm{b}$ up to $\approx~-1~\%$. Under $c$-axis compression (Fig.~\ref{fig:raw-data}~\textbf{D}), the large, negative peak in $\eta_\textrm{c}(T)$ is clearly visible for all $\varepsilon_\textrm{c}$, and the corresponding $T^\star$ moves to higher temperatures with increasing compression. In addition, for $\varepsilon_\textrm{c}~\lesssim~-0.62~\%$, a second but smaller anomaly in $\eta_\textrm{c}$ is detected at $T^+>T^\star$. Unlike the peak at $T^\star$, the step-like anomaly at $T^+$ is typical of a mean-field second-order phase transition. $T^+$ rapidly shifts to higher temperatures with increasing compression. The discovery of a second, `new' phase transition is the key result of the study under $c$-axis compression (i.e., when tuning to lower frustration).

Compression along the $b$-axis yields a different behavior of the ECE. Specifically, the large, positive peak in $\eta_\textrm{b}$ at $T^\star$ becomes significantly smaller with increasing compression and eventually changes sign, so that a negative peak is observed for $\varepsilon_\textrm{b}~\lesssim~-0.55~\%$. Importantly, $T^\star$ itself changes only weakly: with compression, it is first slightly suppressed and then increases again, but is not suppressed to zero temperature. This result implies that the zero-strain, spin-gapped ground state remains stable under large, anisotropic $b$-axis strain. As we will discuss further below, the strain-induced sign change of the anomaly in $\eta_\textrm{b}$ is particularly significant and strongly suggests that we have tuned $\kappa$-(ET)$_2$Cu$_2$(CN)$_3$ to, and even beyond, the point of maximum frustration.

\subsubsection*{The temperature-strain phase diagram}

\begin{figure}
\centering
\includegraphics[width=\textwidth]{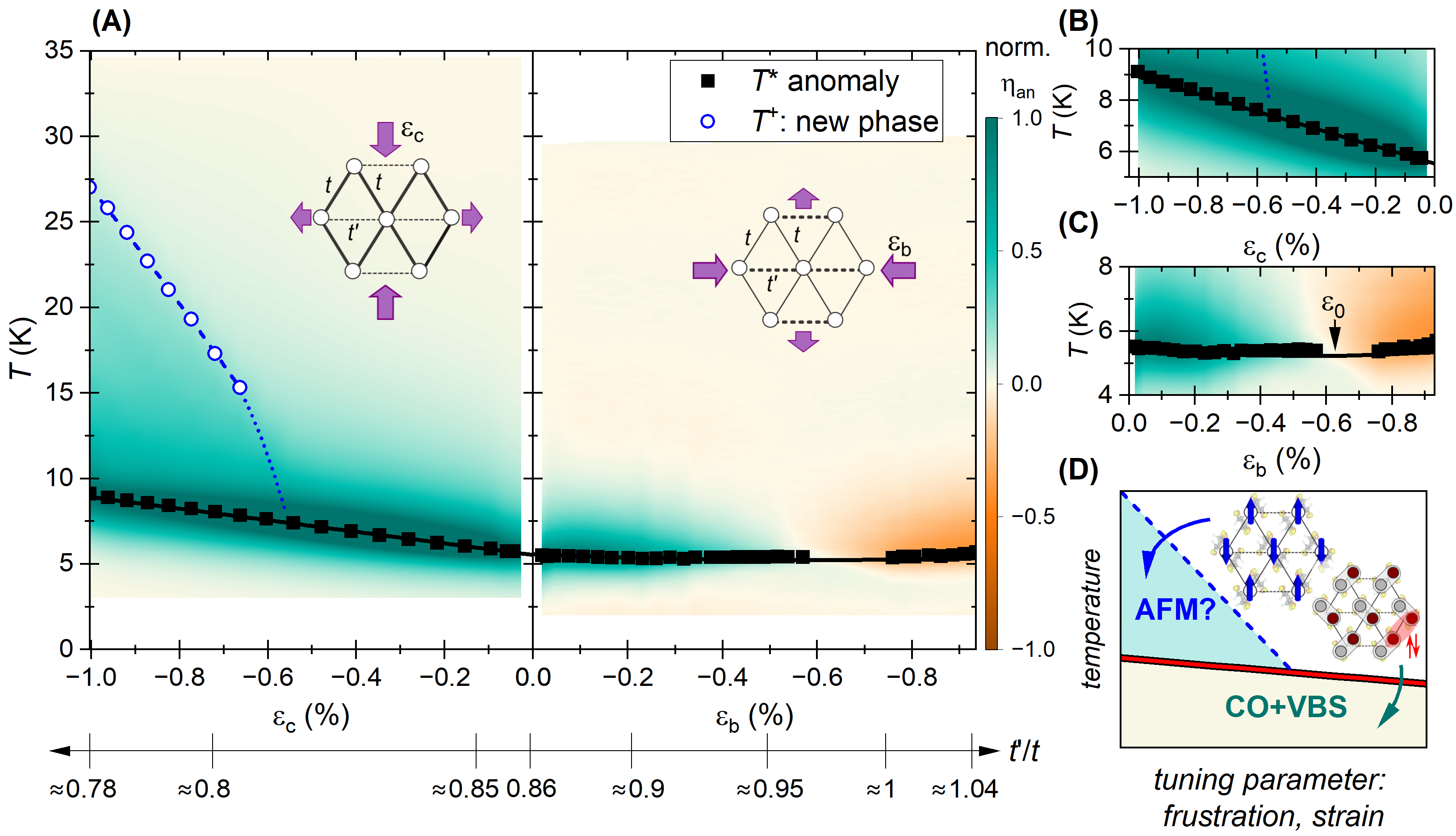} 
\caption{\textbf{Phase diagram of $\kappa$-(ET)$_2$Cu$_2$(CN)$_3$ under variable strain and frustration.} \textbf{(A)} Experimental phase diagram as a function of $\epsilon_b$ and $\epsilon_c$, together with estimates for the changes in the frustration ratio $t^\prime/t$ (see Supplementary Information), showing the large variation of $t^\prime/t$ induced in our experiment. The color plot shows the anomalous contribution to the elastocaloric signal, $\eta_\textrm{an}$, normalized to the $\eta_\textrm{an}$ value at zero strain. The new phase (blue circles) is present at $t^\prime/t~\lesssim~0.82$ and $\varepsilon_\textrm{c}~\lesssim~-0.55~\%$. The blue dotted line corresponds to the extrapolation of the $T^+$-line to lower temperatures. The characteristic temperature $T^\star$ (black squares) adopts a minimum and $\eta_\textrm{an}$ changes sign when $t^\prime/t$ is $\approx~$1.  (\textbf{B}, \textbf{C}) Enlarged view of the low-temperature part of the phase diagram. No transition temperatures were extracted in regions, where the elastocaloric anomaly is small because d$T^\star$/d$\varepsilon~\approx~0$, i.e. close to $\varepsilon_0$ (see black arrow). The solid line in \textbf{C} corresponds to the strain dependence of $T^\star$, calculated from the size of the elastocaloric anomaly under the assumption of a constant change of entropy at the phase transition. This analysis shows that $T^\star$ remains finite across the entire strain range. (\textbf{D}) Proposed phase diagram around the polycritical point at low frustration, based on symmetry considerations and Landau free energy analysis (see Supplementary Information). A second-order antiferromagnetic (AFM) transition line (dashed blue line) meets a first-order line (solid black and red line) at which intra-dimer charge order (CO)~\cite{AbdelJawad10,Matsuura22,Kobayashi20} and valence-bond solid (VBS) order can occur simultaneously. In the CO state, charge is localized towards one of the two ET molecules that form the dimers. Brown (gray) circles indicate charge-rich (charge-poor) ET sites.}
\label{fig:phasediagram}
\end{figure}

The above data allow us to construct a unified phase diagram (Fig.~\ref{fig:phasediagram} \textbf{A}), consisting of the experimental data of the ECE as a function of both $c$- and $b$-axis strains, i.e., under decreasing and increasing $t^\prime/t$. To this end, we consider the anomalous contribution to the experimental $\eta$, which we label $\eta_\textrm{an}$, obtained after subtracting a background contribution from the experimental data (see Supplementary Information). Figure~\ref{fig:phasediagram} \textbf{A} shows a color plot of $\eta_\textrm{an}$, overlaid with the extracted phase transition temperatures $T^\star$ and $T^+$. Before proceeding with a discussion of our results, we use this phase diagram to demonstrate that the observed elastocaloric anomaly at $T^+$ is fully consistent with the expected signature of a thermodynamic phase transition. To rationalize our observations, we use a modified Ehrenfest relation for a second-order phase transition at a critical temperature $T_c$, which is formulated via $\eta$ and $C$ as~\cite{Jerzembeck24}

\begin{equation}
\frac{\Delta \eta}{\eta_h(T_c)} = -\frac{1-\frac{1}{\eta_h (T_c)}\frac{\textnormal{d}T_c}{\textnormal{d}\varepsilon}}{1+\frac{C_h(T_c)}{\Delta C}},
    \label{eq:Ehrenfest2}
\end{equation}

\noindent with $\Delta \eta = \eta (T~\rightarrow T_c^-)-\eta (T~\rightarrow T_c^+)$ and $\Delta C = C (T~\rightarrow T_c^-)-C (T~\rightarrow T_c^+)$ the discontinuities of the elastocaloric effect and the specific heat at $T_c$, respectively, and $\eta_h (T_c)=\eta (T~\rightarrow T_c^+)$ and $C_h (T_c)=C (T~\rightarrow T_c^+)$ being the ECE and the heat capacity of the high-temperature phase, which are related to background contributions, e.g. from phonons. For the $T^+$ transition, the phase diagram in Fig.~\ref{fig:phasediagram}~\textbf{A} shows a large, negative slope, d$T^+$/d$\varepsilon_\textrm{c}$, yet the anomaly in $\eta$ is much smaller in magnitude than the one at the $T^\star$ transition (see Fig.~\ref{fig:raw-data}~\textbf{D}). Following Eq.~\ref{eq:Ehrenfest2}, this observation can be rationalized if $\frac{C_h(T^+)}{\Delta C}$ is large compared to $\frac{C_h(T^\star)}{\Delta C}$. Given that the specific heat of organic charge-transfer salts is dominated by phonon contributions, raising strongly with temperature, it is reasonable that $C_h(T^+\approx~20~$K) is much larger than $C_h(T^\star~\approx~6~$K). Taking the experimental values of d$T^+$/d$\varepsilon_\textrm{c}~\approx~-3500~$K, $\Delta \eta(T^+)~\approx~-23$~K, $\eta_h(T^+)~\approx~-16$~K (see Supplementary Information) and using the ambient-pressure value~\cite{Yamashita08} $C_h(T~=~20~$K$)~\approx~65$~J/(mol$~$K) gives $\Delta C~\approx~0.4$~J/(mol$~$K). This estimate of $\Delta C$ is consistent with the expectations for electronic phase transitions in the $\kappa$-(ET)$_2X$ salts~\cite{Riedl21b,Manna18}. 

Finally, the phase diagram also highlights the sign change of $\eta_\textrm{an}$ under $b$-axis strain. The peak height, $\Delta \eta (T^\star)$, is related to $\frac{\textnormal{d}T^\star}{\textnormal{d}\varepsilon_\textrm{i}}$ (see Supplementary Information). As a result, the behavior of $\eta_\textrm{an}$ under $b$-axis strain indicates that $\frac{\textnormal{d}T^\star}{\textnormal{d}\varepsilon_\textrm{b}}$ is positive at zero strain, then decreases and becomes negative at $\varepsilon_0~\approx~-0.55~\%$, suggesting a large quadratic contribution to $T^\star(\epsilon_b)$. Indeed, our experimental phase diagram is well described by an in-strain quadratic function (see black line in Fig.~\ref{fig:phasediagram} \textbf{a,c}), with a clear minimum in $T^\star(\epsilon_b)$ observed at $\varepsilon_0$. The symmetric behavior of $T^\star$ with strain around $\varepsilon_0$ is a crucial result, suggesting that the system has a symmetry at $\varepsilon_0$ that is broken by strain (see, e.g., the interpretation of ECE data on nematic phase transitions~\cite{Ikeda21}, where the application of small strains breaks tetragonal symmetry). The isotropic triangular lattice possesses such a symmetry, thus it is natural to associate $\varepsilon_0$ with the strain at which $\kappa$-(ET)$_2$Cu$_2$(CN)$_3$ realizes the isotropic triangular lattice. 

The unified phase diagram in Fig.~\ref{fig:phasediagram}~\textbf{A} links the central experimental findings of the present study to the corresponding $t^\prime/t$ values, which we estimate based on the results of density-functional theory (DFT) calculations for different lattice structures of $\kappa$-(ET)$_2$Cu$_2$(CN)$_3$ in Ref.~\citen{Jeschke12} (see Supplementary Information). First, the new $T^+$ phase, which we observed under compressive $\varepsilon_\textrm{c}$ strain, occurs at $t^\prime/t~\lesssim~0.82$. Second, the spin-gapped ground state below $T^\star$ is stable for the entire range of strains and frustrations studied ($0.78~\lesssim~t^\prime/t~\lesssim~1.04$). Most importantly, the $t^\prime/t$ estimates confirm that the spin gap is stable at maximum frustration ($t^\prime/t~\approx~1$). 

\subsection*{Discussion}

\noindent Our study on the frustrated Mott insulator $\kappa$-(ET)$_2$Cu$_2$(CN)$_3$ demonstrates that anisotropic strain is an exceptionally effective tuning parameter for probing the phase diagram of frustrated quantum magnets. By continuously varying the ratio of hopping parameters $t'/t$ of $\kappa$-(ET)$_2$Cu$_2$(CN)$_3$ over a wide range as large as 25\,\%, we uncover a rich landscape of competing low-temperature orders. These findings provide fresh insights into the physics of the $\kappa$-(ET)$_2X$ family, facilitate comparisons with theoretical models and, more generally, open up new experimental avenues for investigating a wide range of geometrically frustrated quantum magnets.

First, we focus on the new insights into the physics of $\kappa$-(ET)$_2$Cu$_2$(CN)$_3$ from the phase diagram. A key finding is the identification of a polycritical point at $(\varepsilon_\mathrm{c}, T) \approx (-0.55\%, 7.5\ \mathrm{K})$, where the phase boundaries of the $T^\star$ and $T^+$ phases meet. It challenges usual expectations from Landau theory (schematically depicted in Fig.~\ref{fig:overview}~\textbf{B}), since the $T^\star$ transition line remains unaffected by the appearance of the $T^+$ phase. This discrepancy can only be reconciled with Landau theory if the $T^\star$ transition is first order across the entire range of frustration (see Supplementary Information). The symmetric peak in $\eta_\textrm{i}$ at $T^\star$ supports the first-order nature of $T^\star$ (see Supplementary Information), as does the recent observation of sudden phonon mode changes at $T^\star$~\cite{Matsuura22}. Together, these results suggest a fundamental shift in our understanding of the long-debated '6~K anomaly' in $\kappa$-(ET)$_2$Cu$_2$(CN)$_3$.

In the following, we argue based on symmetry considerations that the $T^\star$ transition must be first order if it arises from the coupling of two distinct order parameters: a spin-gapped valence-bond solid (VBS) and an intra-dimer charge order (CO). The latter has recently added a new twist to the discussion of the physics of the family of organic $\kappa$-(ET)$_2X$ salts, since the coupling of intra-dimer charge degrees of freedom with spin degrees of freedom can lead to new forms of electronically-driven multiferroicity~\cite{Lunkenheimer12,Gati17,Hassan18,Lang25}. In $\kappa$-(ET)$_2$Cu$_2$(CN)$_3$, a large set of recent experiments suggest the presence of intra-dimer CO below $T^\star$\cite{AbdelJawad10,Manna10,Matsuura22,Kobayashi20,Liebmann24}. It is important to note that VBS and CO break distinct symmetries. While the CO state preserves both $2_1$ screw symmetry and lattice translation, the various possible dimer patterns for the VBS must break a subset of these symmetries. Their simultaneous onset at $T^\star$ requires a strong coupling, which enforces a first-order transition (see Supplementary Information). Identifying the `6~K-anomaly' as a first-order transition here provides important support for the notion that the weakly Mott insulating $\kappa$-(ET)$_2X$ salts display a rich cooperative behavior of spin and charge degrees of freedom, that may lead to novel functionalities.

The so far unidentified $T^+$ phase emerges at lower frustration, below a critical ratio of $t^\prime/t~\approx~0.8$. A natural candidate for this $T^+$ phase is AFM order, consistent with numerical studies of the triangular-lattice Hubbard model \cite{Szasz21}. The $\Delta C$ estimate for the $T^+$ transition is consistent with a magnetic phase transition~\cite{Riedl21b,Manna18}: for the related, less frustrated compound $\kappa$-(ET)$_2$Cu[N(CN)$_2$]Cl with $t^\prime/t~\approx~0.44$, a negligible $\Delta C$ below experimental resolution\cite{Yamashita10} at the AFM transition has been reported. Since AFM order breaks yet other symmetries than VBS or VBS+CO, it competes with the $T^\star$ phase, in agreement with our simulations of the polycritical point (see Fig.~\ref{fig:phasediagram}~\textbf{D}). Alternatively, the $T^+$ phase may represent a distinct intra-dimer charge-ordered state, which would also compete with the VBS+CO state.

We conclude that anisotropic strain offers a new opportunity for tuning emergent cooperative charge-spin phenomena in frustrated $\kappa$-(ET)$_2X$. Our advancements suggest that microscopic magnetic, dielectric, or optical measurements under large anisotropic strains can be used to disentangle competing charge-ordered and magnetic ground states. In particular, determining whether the $T^+$ phase breaks spin or charge symmetries will be the key next step for understanding the interplay between these intertwined orders.

More broadly, our phase diagram also establishes a new benchmark for comparison with theoretical studies of correlated electrons on frustrated lattices. A considerable body of work is devoted to uncovering the rich landscape of phases predicted for the triangular-lattice Hubbard model at intermediate interaction strength, $U/t$, near the Mott metal-insulator transition and across a broad range of frustration, $t'/t$. Resolving the phase diagram in this parameter space is notoriously difficult, as competing ground states often lie extremely close in energy and challenge the most advanced numerical methods. The regime we access by tuning $t^\prime/t$ by roughly 25\% coincides with the parameter space where several candidate phases have been proposed. Thus, our experimentally determined phase diagram as a function of frustration offers an important reference point for testing how reliably state-of-the-art calculations capture the physics of real materials. For example, Szasz \textit{et al.}~\cite{Szasz21} report that AFM order emerges robustly for $t'/t~\lesssim 0.8$, independent of boundary conditions. As noted above, this value agrees well with the appearance of the $T^+$ phase in our measurements. Near the regime of maximal frustration, competing ground states such as gapped chiral spin liquids and valence-bond solids have been proposed~\cite{Szasz20,Wietek21,Wietek17,Wietek24}, but their stability in the presence of hopping anisotropy ($t'/t~\neq~1$) remains poorly understood. Our result that the $T^\star$ phase is robust under tunable strain may serve as a critical input for resolving these open questions.

Finally, our approach introduces a broadly applicable framework for accessing otherwise hidden regions of phase space in frustrated quantum materials. The combination of controlled strain tuning with the elastocaloric probe enables the discovery of emergent phases driven by electron-lattice interactions and direct tests of theoretical models. Looking ahead, applying this methodology to other frustrated systems, such as kagome magnets or delafossite triangular-lattice spin liquids~\cite{Scheie24}, promises to uncover a deeper understanding of the interplay between geometry and correlations in quantum matter.

\subsection*{Materials and Methods}

\paragraph*{Crystal growth and sample preparation -} The crystals of $\kappa$-(ET)$_2$Cu$_2$(CN)$_3$ were grown using electrochemical crystallization~\cite{Geiser91}. The crystals were typically 1-2~mm long in each in-plane direction, with a thickness of $\approx~100~\mu$m. Samples were mounted on the strain cell either in their as-grown form or after being polished with a Plasma Focused Ion Beam. Details of the sample preparation procedure are described in the Supplementary Information, which also includes photographs of the preparation steps.

\paragraph*{Strain application -} Strain was applied either along the crystallographic $b$- or $c$-axis \textit{in situ} at low temperature using a piezoactuator-driven uniaxial pressure cell, similar in design to the one reported in Ref.~\citen{Barber19}. To ease sample preparation, samples were mounted across a gap between a fixed and a moving part of a sample carrier, which were subsequently mounted onto the pressure cell. The sample carrier is designed such that only compression can be applied to our sample to determine the zero-strain state with high precision (see Supplementary Information for details).

\paragraph*{Elastocaloric effect measurements -} To measure the a.c. elastocaloric effect~\cite{Ikeda19} as a function of tuning strain, small a.c. voltages, with frequencies between 40 and 100 Hz, were used to modulate the d.c. voltages on the piezoelectric actuators, which provide the d.c. tuning strain (see Supplementary Information for details on determining the optimal measurement frequency). As a result, the a.c. strain amplitude $\Delta \varepsilon~~\approx~3~\cdot~10^{-5}$. The resulting temperature oscillation, $\Delta T$, was measured using a chromel–AuFe$_{0.07\%}$ thermocouple that was attached to the sample. The thermocouple voltage was amplified by a low-temperature transformer and a room-temperature preamplifier and recorded using a Lock-In amplifier.

\noindent \textbf{Acknowledgments:}
We acknowledge useful discussions with Stuart Brown, Roser Valent\'i and Andy Mackenzie. 
\textbf{Funding:}
Financial support by the Max Planck Society is gratefully acknowledged. This study was supported through the Deutsche Forschungsgemeinschaft (DFG, German Research Foundation) through TRR 288—422213477 (projects A13 and A06) and through SFB 1143 on Correlated Magnetism (Project No. 247310070 - project C09) as well as by the Japan Society for the Promotion of Science KAKENHI Grant No. 23K25811.
\textbf{Author contributions:}
 E.G. and M.L. designed and supervised the project. F.L., Y.S., Y.A. and E.G. conducted and analyzed the experiments. T.S. and N.Y. grew the single crystals used in this study. S.M.W. developed the theoretical models. E.G. wrote the manuscript with contributions from all authors.
\textbf{Competing interests:}
There are no competing interests to declare. \\
\textbf{Data and materials availability:}
The data that support the findings of this study will be made openly available in the Max Planck Digital Library upon publication.

\newpage

\bibliographystyle{apsrev}
\bibliography{main-kappa-CuCN-arXiv} 

\newpage
\subsection*{Supplementary Text}

\renewcommand{\thefigure}{S\arabic{figure}}
\setcounter{figure}{0}

\renewcommand{\thetable}{S\arabic{table}}
\setcounter{table}{0}

\subsubsection*{\underline{\small Details of Experimental Methods}}

\noindent \textit{Sample preparation - } Experiments were conducted on in total six samples: four (two) were mounted in such a way that uniaxial pressure was applied along the $c$ ($b$) axis (see \ref{tb:SI-samplesoverview}). The key findings discussed in the main text, including the sign change of $\eta_\textrm{an}$ under $b$-axis strain and the emergence of the new phase under $c$-axis strain, were verified in the two samples (per direction) that withstood high compression.

\begin{table}[h!]
\centering
\caption{\textbf{Overview of samples of $\kappa$-(ET)$_2$Cu$_2$(CN)$_3$ studied in the present work.} The crystallographic direction of applied pressure is indicated in the second column. (*) marks the samples on which the data in the main text was taken. \\}
\label{tb:SI-samplesoverview}
\begin{tabular}{ |c|c| } 
 \hline
 sample no. & pressure direction  \\ \hline
 \#1 & $c$  \\ \hline
 \#2 (*) & $b$  \\ \hline
 \#3 & $c$  \\ \hline
 \#5 & $b$  \\ \hline
 \#6 (*) & $c$  \\ \hline
 \#10 & $c$  \\ \hline
\end{tabular}

\end{table}

\begin{figure}[h!]
\centering
\includegraphics[width=0.7\columnwidth]{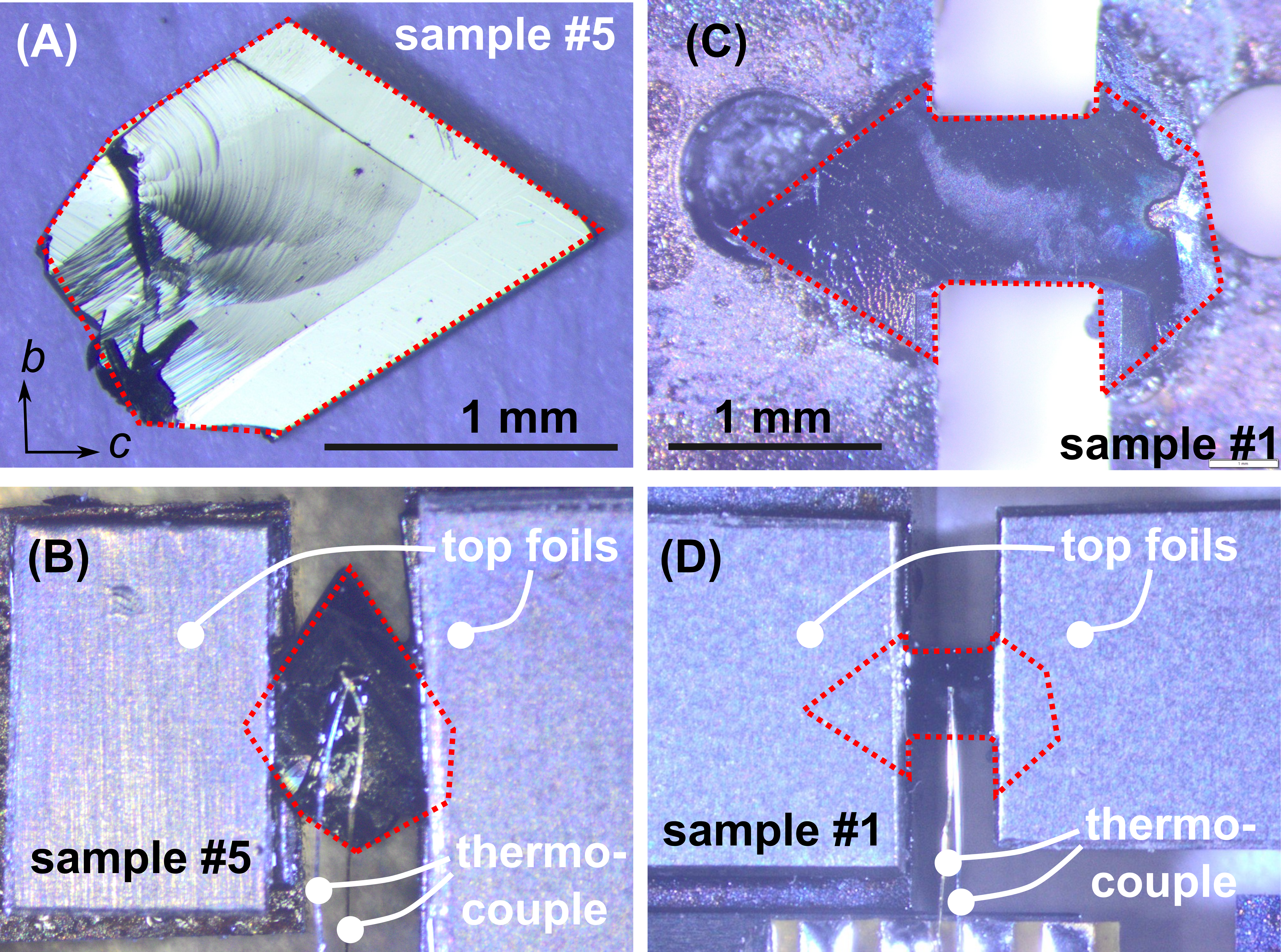} 
\caption{\textbf{Mounting of $\kappa$-(ET)$_2$Cu$_2$(CN)$_3$ crystals for experiments under uniaxial pressure and anisotropic strains.} (a) As-grown crystal of $\kappa$-(ET)$_2$Cu$_2$(CN)$_3$ (sample \#5). (b) The as-grown crystal (\#5) mounted across a gap on the sample carrier, with side foils (not shown) and top foils on the ends to ensure uniform strain transmission throughout the thickness of the sample. A thermocouple is attached to the middle of the sample. (c) $\kappa$-(ET)$_2$Cu$_2$(CN)$_3$ crystal (sample \#1) polished and shaped with a Plasma Focused Ion Beam (PFIB), yielding a well-defined cross-section across the gap that enhances strain uniformity in the strained section and enables precise pressure determination. (d) $\kappa$-(ET)$_2$Cu$_2$(CN)$_3$ crystal (sample \#1) with top foils and thermocouple attached. The crystal shape is indicated by the red dotted line in each subpanel.}
\label{fig:SI-mounting}
\end{figure}

We mounted the crystals either in their as-grown form or polished them using a Plasma Focused Ion Beam (PFIB) for reasons detailed below. Figure~\ref{fig:SI-mounting}~(a) shows a typical example of a crystal in its as-grown state, with well-defined facets corresponding to (1~1~0) planes (sample \#5). Initially, each crystal is mounted along either the $b$- or $c$-axis across an $\approx~$800$~\mu$m gap on the sample carrier, as illustrated for $b$-axis sample \#5 in fig.~\ref{fig:SI-mounting}~(b). The gap size is precisely adjusted \textit{in situ} during the experiment using a piezoactuator-driven uniaxial-pressure cell to compress the sample. The sample ends are secured to the carrier using Stycast 1266 (two-component epoxy) and encapsulated with side and top foils to ensure uniform strain transmission throughout the sample’s thickness. A thermocouple is attached to the center of the sample with a small drop of Stycast 1266 to measure temperature changes induced by an a.c. strain.

Applying uniaxial pressure to an as-grown sample introduces some uncertainty in pressure determination due to its non-uniform cross-section. Additionally, strain relaxation can occur across the sample's width, particularly near the free-standing edges of the crystal. Although the section of the sample probed by the thermocouple can be controlled by adjusting the measurement frequency (see fig.~\ref{fig:SI-frequencysweep}), we aimed to improve pressure homogeneity and accuracy by shaping the crystals to have a uniform cross-section. Conventional hand-polishing tools are unsuitable for these delicate crystals, so we employed a Xe PFIB (Plasma Focused Ion Beam) to polish the samples with a 0.2 $\mu$A current, creating an $\approx~$750$~\mu$m long section with uniform width (as shown in fig.~\ref{fig:SI-mounting}~(c) for crystal \#1 post-polishing). The sample ends attached to the carrier remained unpolished. Similar to the as-grown samples, the PFIB-polished crystals had their ends encapsulated with side and top foils, and a thermocouple was affixed to the sample (see fig.~\ref{fig:SI-mounting}~(d)). 

\noindent \textit{Uniaxial pressure cell - } The piezoactuator-driven uniaxial pressure cell used is similar to that described in Ref.~\citen{Barber19} and is equipped with two capacitive sensors to measure the sample’s displacement and applied force. It is important to note that the displacement sensor is mounted below the sample, so the recorded displacement includes contributions from mounting glues etc.. Therefore, the measured displacement must be corrected to accurately determine the sample strain (for further details, see Ref.~\citen{Hicks14}). The force is determined based on the deformation of flexures with a calibrated spring constant within the cell.

\begin{figure}[h!]
\centering
\includegraphics[width=0.7\columnwidth]{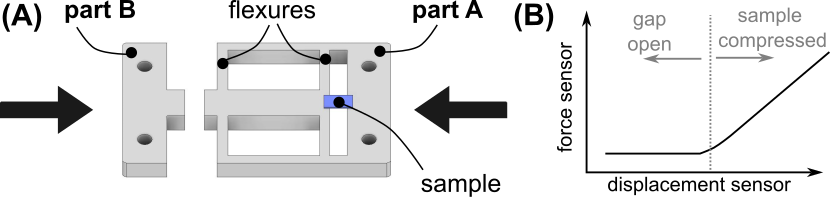} 
\caption{\textbf{Sample carrier used for the uniaxial pressure study. }(a) Schematic illustration of the two-part sample carrier used in this study. The carrier consists of Parts A and B, with the sample mounted across the gap between the movable section (guided by flexures) and the fixed section of Part A. Arrows indicate the direction of force applied by the uniaxial pressure cell. Compression of the sample occurs only when the pressure cell closes the mechanical contact between Parts A and B. (b) Schematic force-displacement curve as measured by the cell's force and displacement sensors when using a two-part carrier. The abrupt change in slope at the gray dotted line indicates the point where mechanical contact between Parts A and B is achieved as displacement increases.}
\label{fig:SI-samplecarrier}
\end{figure}

\begin{figure}[h!]
\centering
\includegraphics[width=0.6\columnwidth]{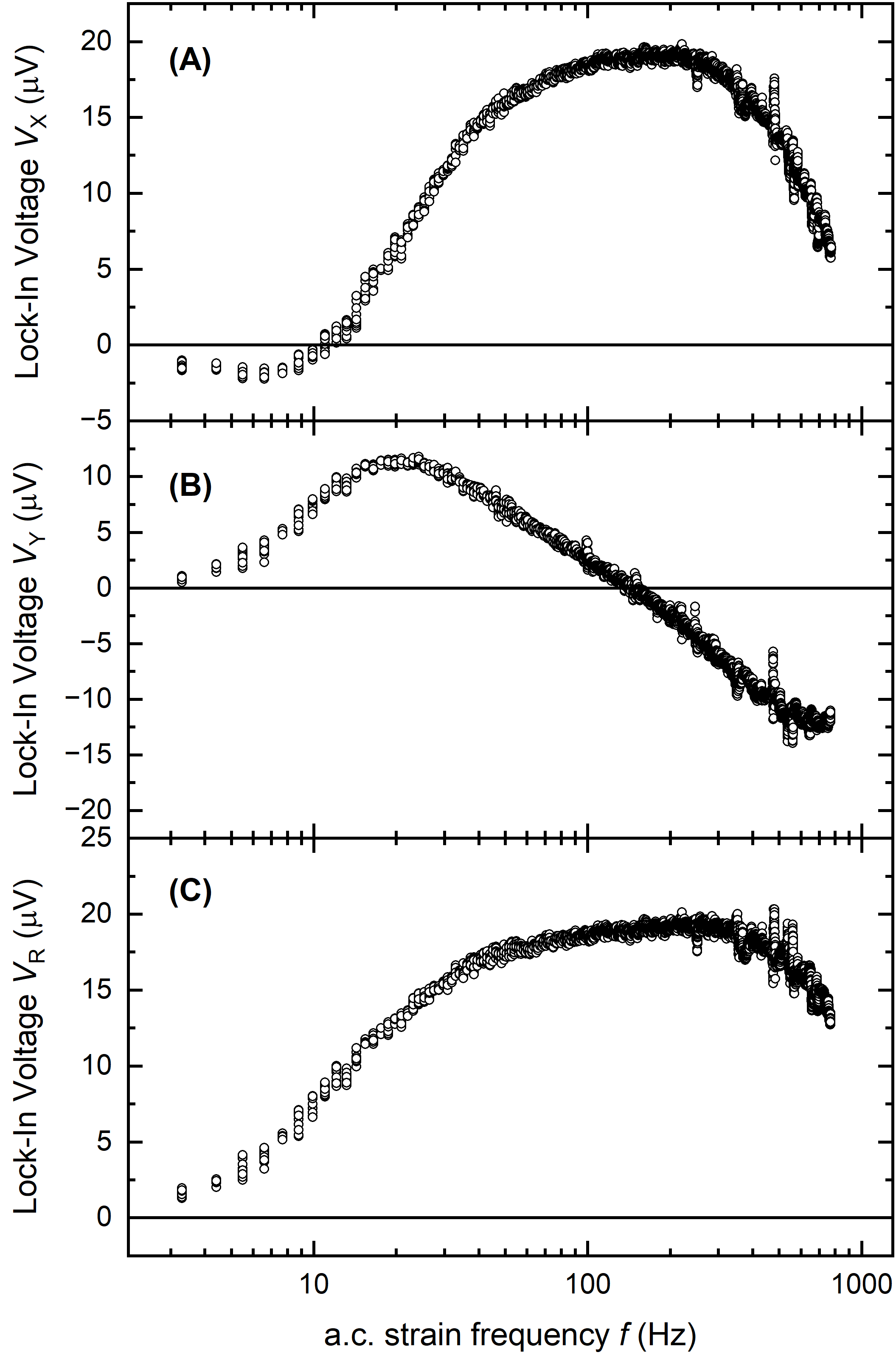} 
\caption{\textbf{Determination of the optimal a.c. measurement frequency for elastocaloric experiments via a frequency sweep.} The in-phase part (a), out-of-phase part (b), and magnitude (c) of the thermocouple’s a.c. voltage, recorded by a Lock-in amplifier, are measured as a function of frequency (shown here for sample \#2 at a temperature of 6~K and near zero strain). }
\label{fig:SI-frequencysweep}
\end{figure}

\noindent \textit{Sample carrier design - } Here, we describe the sample carrier used in this study, which is made out of titanium, like the main body of the cell. While its main function is to enable easy sample preparation and exchange, the two-part design of this carrier also offers the advantage of precise, \textit{in situ} control and determination of the zero-force state at low temperatures (see also Ref.~\citen{Jerzembeck22}). The operating principle of the two-part carrier is illustrated in fig.~\ref{fig:SI-samplecarrier}(a). It consists of two main parts, labeled as Part A and Part B. Part A is further divided into a movable section, guided by flexures, and a fixed section, with the sample mounted across the gap between these sections of Part A.

To apply pressure to the sample, Part A is brought into mechanical contact with Part B, which is connected to the pressure cell. The closure of the mechanical contact is detected through force-displacement curves recorded by capacitive force and displacement sensors within the cell (see fig.~\ref{fig:SI-samplecarrier}(b)). When displacement changes without a corresponding change in force, the contact remains open. Conversely, if force varies with displacement, the sample is under compression. By adjusting this contact \textit{in situ} at any temperature, we can determine capacitance readings that indicate zero force and zero displacement — something typically challenging in conventional sample mounting methods (see Ref.~\citen{Hicks14}) due to differential thermal expansion between the sample and cell. However, with this two-part carrier, the sample cannot experience tensile strain. As a consequence, the elastocaloric effect cannot be measured at exactly zero strain or force but requires a small d.c. tuning strain, as an exact zero strain would cause the a.c. strain modulation to open the gap.

\noindent \textit{Elastocaloric measurement frequency - } To achieve quasi-adiabatic measurement conditions, the measurement frequency, $f$, must be selected such that $1/\tau_1 \ll 2 \pi f \ll 1/\tau_2$, where $\tau_1$ represents the relaxation time between the sample and the bath, and $\tau_2$ is the relaxation time between the sample and the thermocouple. These relaxation times depend on material properties, such as thermal conductivity and specific heat, and vary across different sample mountings due to differing amounts of glue and other factors. Therefore, the optimal measurement frequency must be determined for each experiment. An example frequency sweep for $\kappa$-(ET)$_2$Cu$_2$(CN)$_3$ is shown in fig.~\ref{fig:SI-frequencysweep}. When the quasi-adiabatic condition is met, the in-phase component of the thermocouple’s a.c. voltage, $V_\textrm{X}$, as well as the magnitude, $V_\textrm{R}$, reach their maximum values, while the out-of-phase component, $V_\textrm{Y}$, remains small (ideally zero). In this example, this condition is satisfied within a frequency range centered around 130~Hz. At much lower or much higher frequencies, the generated heat either dissipates into the bath or does not fully reach the thermocouple, resulting in significantly reduced values for $V_\textrm{X}$ and $V_\textrm{R}$ compared to their maximal values.

\clearpage

\subsection*{\underline{\small Thermodynamic consistency with ambient-pressure literature data}}

\noindent \textit{Calibration of the ECE signal - }In the following, we show that our elastocaloric effect (ECE) data is consistent with ambient-pressure thermodynamic data, reported in literature~\cite{Manna10,Manna18,Yamashita08}. In general, the ECE (see Eq. 1 in the main text) is related to the directional-dependent thermal expansion coefficient, $\alpha_\textrm{i}(T) = \frac{1}{V} \left( \frac{\partial S}{\partial p_i} \right)_T$, and the specific heat, $C_p(T) = T \left( \frac{\partial S}{\partial T} \right)_p~\approx C_V (T)$. Note that we use a convention for pressure, where compression is indicated by a negative sign. It follows that

\begin{equation}
    \left (\frac{\partial T}{\partial \varepsilon_\textrm{i}}\right)_S = -V_\textrm{mol} Y_i T \frac{\alpha_\textrm{i}}{C_\textrm{mol}},
    \label{eq:thermodynamics}
\end{equation}

\noindent with the Young's modulus $Y_i~=~\frac{\textnormal{d} p_i}{\textnormal{d} \varepsilon_\textrm{i} }$ and $C_\textrm{mol}~=~C_V/V_\textrm{mol}$ the molar heat capacity.

\begin{figure}[h!]
\centering
\includegraphics[width=0.5\columnwidth]{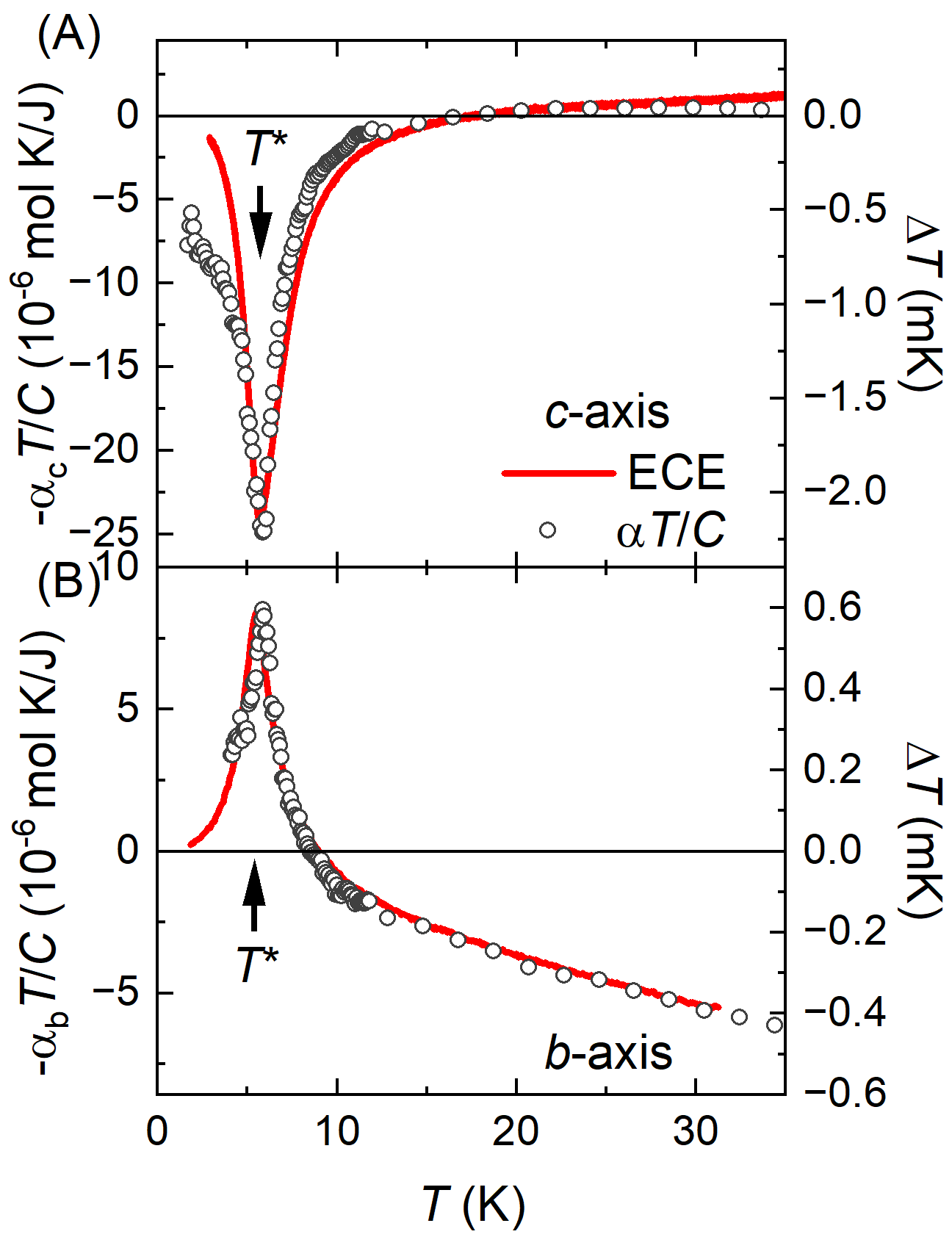} 
\caption{\textbf{Comparison of different thermodynamic data sets, taken on $\kappa$-(ET)$_2$Cu$_2$(CN)$_3$ close to ambient pressure.} In (a) ((b)), the left axis corresponds to $-T\alpha_\textrm{i}/C$ for $i~=~c$ (a) and $i~=~b$ (b) and the right axis corresponds to the measured temperature amplitude induced by a small strain along the $b$-direction in sample \#2 ($c$-direction in sample \#6). Literature data for $\alpha_\textrm{i}(T)$ with $i~=b,c$ and the specific heat $C(T)$ were taken from Refs.~\cite{Manna10,Manna18,Yamashita08}.}
\label{fig:SI-Grüneisen}
\end{figure}

Since we measure at low temperatures, where changes in the elastic moduli are typically very small, the temperature dependence of $Y_i$ can be neglected to a good approximation. Consequently, Eq.~\ref{eq:thermodynamics} suggests that the temperature oscillations measured in our ECE experiments, $ \Delta T $, should be proportional to $ -T \frac{\alpha_\textrm{i}(T)}{C(T)} $. We present this comparison for $\Delta~T$ in fig.~\ref{fig:SI-Grüneisen} (see also Fig.~2 of the main text). It is important to note that in literature some variations in the absolute values of $ \alpha_\textrm{i} $, particularly around the `6~K anomaly' are reported. For our analysis, we used the most recent $ \alpha_\textrm{i} $ data from Ref.~\citen{Manna18}, which reports the largest anomalies near 6~K. 

This thermodynamic analysis allows us to scale our ECE data, as the measured $ \Delta T $ is typically smaller than the actual $ \Delta T $ due to a small degree of non-adiabaticity even at the optimal measurement frequency (see Ref.~\citen{Ikeda19}). By using our estimated Young's modulus (see fig.~\ref{fig:SI-youngsmodulus}), we can calculate the expected $ \eta_\textrm{i} := (\Delta T / \Delta \varepsilon_\textrm{i})_S $ at zero strain. The experimentally-determined value of $ \eta_\textrm{i}$ will be smaller than the expectation based on ambient-pressure thermodynamics not only due to the reduced $\Delta T$, but also due to necessary corrections of $\Delta \epsilon$. Thus, we use the ambient-pressure thermodynamic data to obtain a calibration factor that allows us to convert our measured $ \Delta T $ values into $ \eta $ values. This calibration factor was applied to all data in the main text.

\clearpage

\subsection*{\underline{\small Estimation of the elastic moduli of $\kappa$-(ET)$_2$Cu$_2$(CN)$_3$}}
\label{sec:youngsmod}

We provide here an estimate of the bulk and Young's modulus of $\kappa$-(ET)$_2$Cu$_2$(CN)$_3$, derived from both literature data and our own measurements. Unfortunately, there is no structural data available for $\kappa$-(ET)$_2$Cu$_2$(CN)$_3$ under hydrostatic or uniaxial pressure. Therefore, we refer to existing literature on the structure of the related compound $\kappa$-(ET)$_2$Cu(NCS)$_2$~\cite{Rahal97} and the one-dimensional organic charge-transfer salt (TMTTF)$_2$PF$_6$ under hydrostatic pressure~\cite{Rose13} for an estimate. Organic charge-transfer salts are generally very soft, meaning they exhibit high compressibility (or low moduli). The bulk modulus, $K$, of $\kappa$-(ET)$_2$Cu(NCS)$_2$ is 150 kbar~\cite{Rahal97}, while that of (TMTTF)$_2$PF$_6$ is 100 kbar~\cite{Rose13}. The bulk modulus, $K$, relates to the Young's modulus, $Y$, via Poisson’s ratio, $\nu$, as $Y = 3 K (1 - 2 \nu)$ (assuming an isotropic material for simplicity). For a typical Poisson ratio of $\nu \approx 1/3$, we find that $Y \approx K \approx 100$ kbar. Additionally, $\textnormal{d} p_\textrm{hydro}/(\textnormal{d} b/b)$ and $\textnormal{d} p_\textrm{hydro}/(\textnormal{d} c/c)$ (where $b$ and $c$ are the two in-plane axes) exhibit minimal anisotropy in $\kappa$-(ET)$_2$Cu(NCS)$_2$.

\begin{figure}[h!]
\centering
\includegraphics[width=.7\columnwidth]{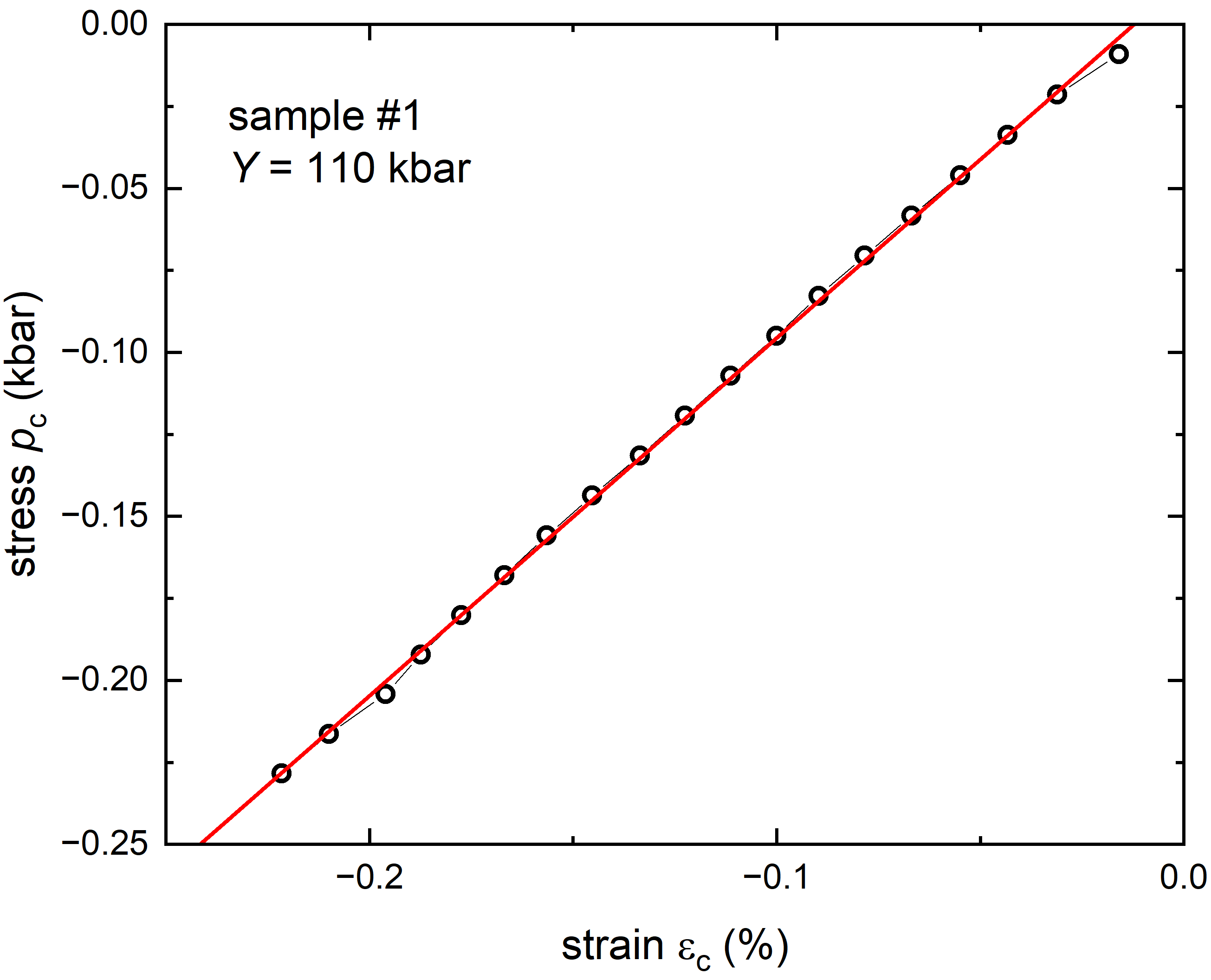} 
\caption{\textbf{Estimation of the Young's modulus of $\kappa$-(ET)$_2$Cu$_2$(CN)$_3$}. The Young's modulus of sample \#1 is estimated using the force and displacement sensors installed in the uniaxial pressure cell.}
\label{fig:SI-youngsmodulus}
\end{figure}

We can also estimate the Young's modulus of $\kappa$-(ET)$_2$Cu$_2$(CN)$_3$ along the $b$- and $c$-axes from our own data, as we measure both the applied force and the resulting displacement using capacitive sensors in our cell~\cite{Noad23}. An example of stress-strain ($\sigma$-$\varepsilon$) data collected from sample \#1 is shown in fig.~\ref{fig:SI-youngsmodulus}. While stress can be calculated with high precision when the sample’s cross-section is well-defined, the measured displacement includes contributions from both the mounting glue and the sample itself. Therefore, to determine the sample strain $\epsilon = \Delta l / l$ (where $l$ is the sample length), the measured displacement, $\Delta d$, must be corrected. This correction can be made using estimates from finite-element simulations (see, e.g., Ref.~\citen{Hicks14}). It is important to note that this correction introduces the main source of error in our estimates of $Y$. After applying this correction to obtain the sample strain, we estimate $Y_c=\frac{\textnormal{d}p_\textrm{c}}{\textnormal{d}\varepsilon_\textrm{c}} \approx 110$ kbar (see fig.~\ref{fig:SI-youngsmodulus}).

\clearpage
\subsection*{\underline{\small Comparison between increasing and decreasing strain runs}}

\begin{figure}[h!]
\centering
\includegraphics[width=.5\columnwidth]{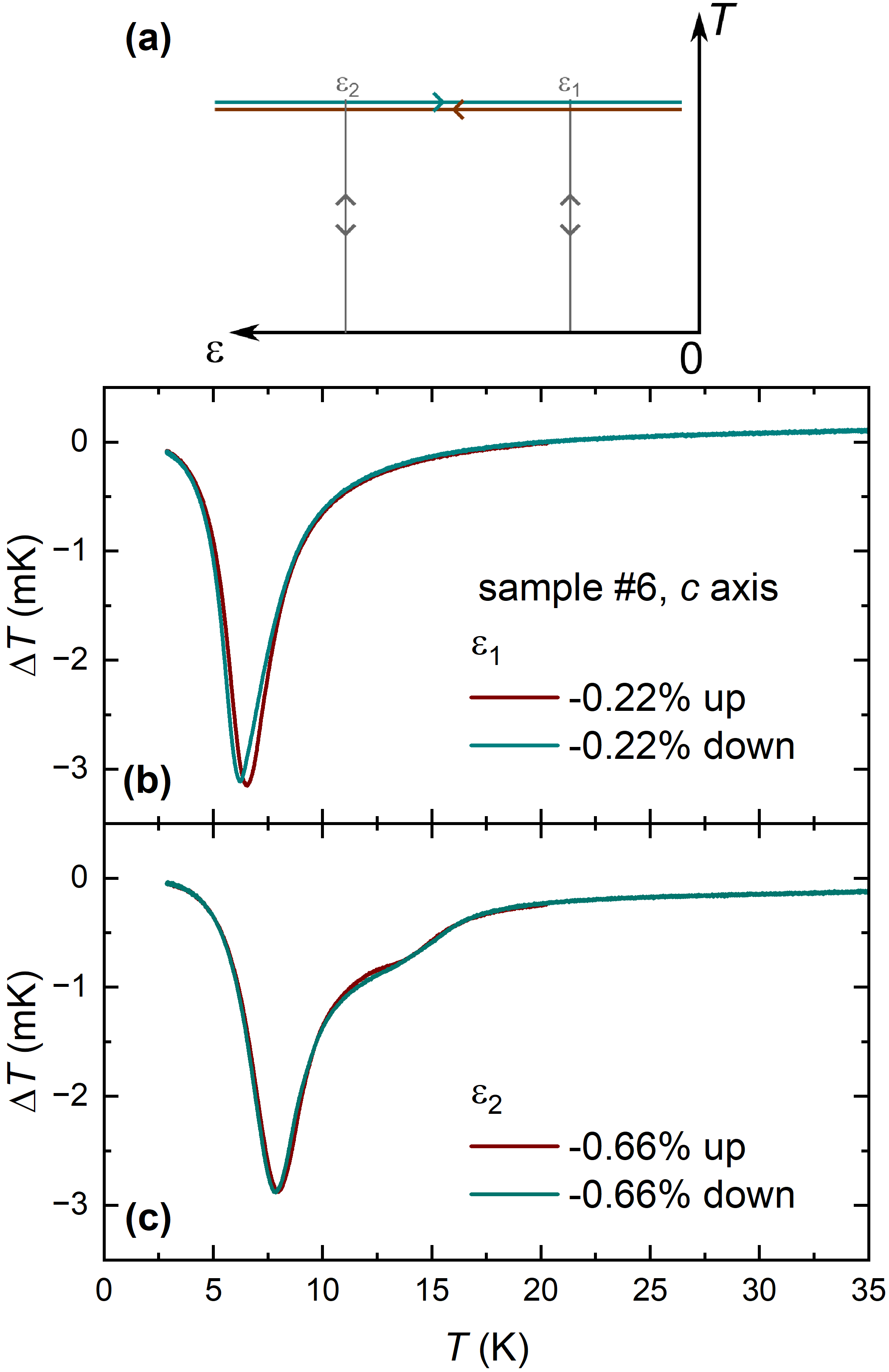} 
\caption{\textbf{Elastocaloric data upon increasing and decreasing pressure. }Comparison of elastocaloric temperature oscillation, $\Delta T$, as a function of temperature, $T$, from runs with increasing and decreasing compression (taken on sample \#6); (a) Schematic illustration of the strain-temperature phase diagram paths. Strain was either increased or decreased at  $T \approx 35~$K, with temperature sweeps conducted at each strain; (b,c) $\Delta T$ vs. $T$ for increasing compression ("up") and decreasing compression ("down") runs at $\varepsilon_1 \approx -0.22\%$ (b) and $\varepsilon_1 \approx -0.66\%$ (c).}
\label{fig:SI-pressurehistory}
\end{figure}

We demonstrate below that the observed behavior of $\kappa$-(ET)$_2$Cu$_2$(CN)$_3$ under large strain is not due to plastic deformation and is entirely reversible as strain is reduced. To show this, fig.~\ref{fig:SI-pressurehistory} presents data from sample \#6 during both increasing and decreasing compression. This sample was strained along the $c$-axis, resulting in significant modifications to the phase diagram, including the emergence of the 'new phase'. In fig.~\ref{fig:SI-pressurehistory}~(a), a schematic of our measurement protocol is shown. Strain was applied at a constant temperature of 35~K, followed by a temperature sweep down to the lowest temperatures and back to 35~K. Compression was initially increased incrementally to a maximum of approximately 1~\%, then decreased stepwise. Figures~\ref{fig:SI-pressurehistory}~(b) and (c) display data from both the increasing and decreasing compression runs at $\varepsilon_1~\approx~-0.22~\%$ (b) and $\varepsilon_1~\approx~-0.66~\%$ (c). The data from both runs show strong agreement within experimental uncertainty, with the new transition at $T^+$, clearly visible at $\varepsilon_2$, vanishing as compression is reduced.

\clearpage
\subsection*{\underline{\small Non-critical and anomalous contributions to the elastocaloric effect}}

\begin{figure}[h!]
\centering
\includegraphics[width=.8\columnwidth]{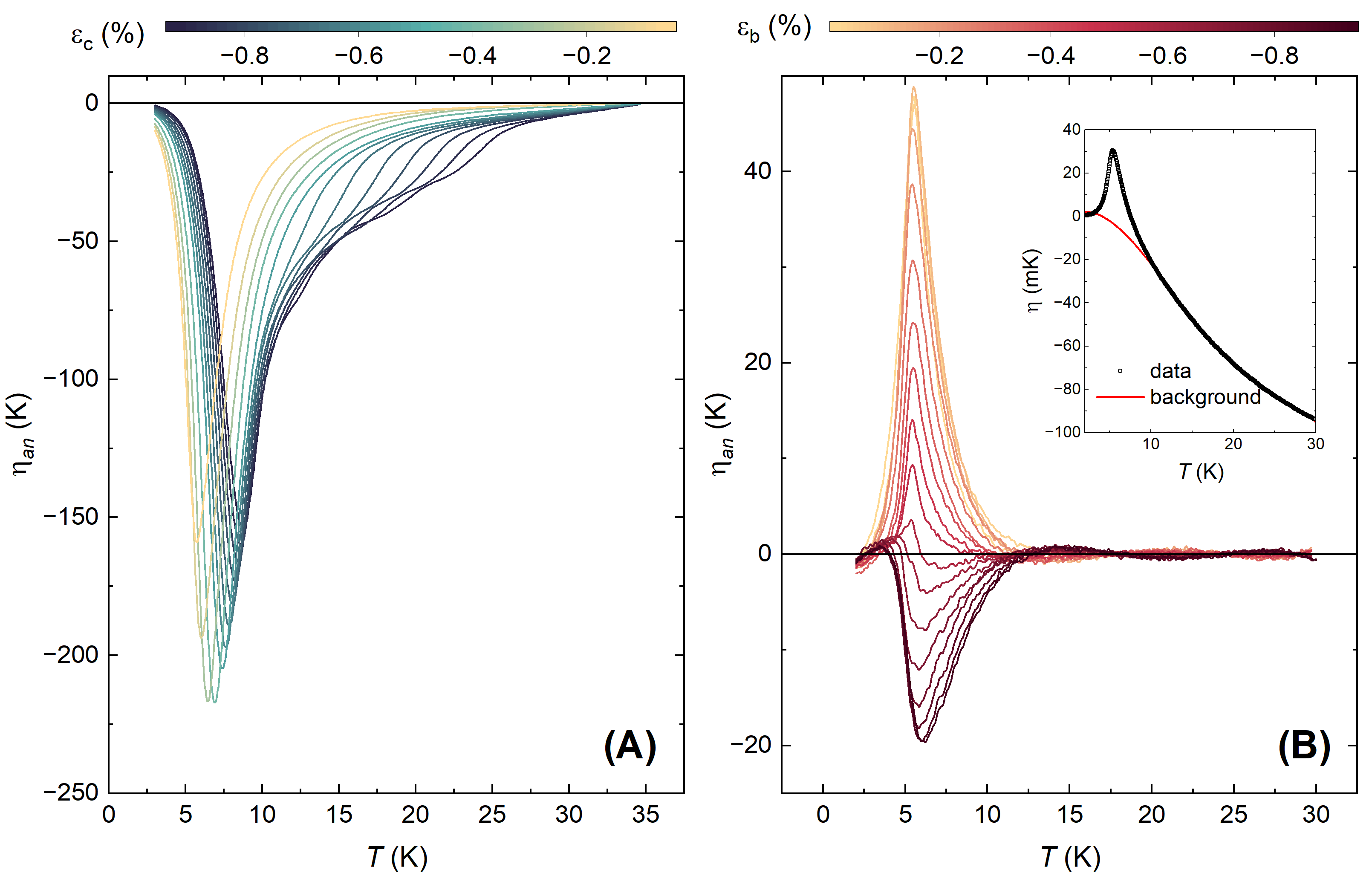} 
\caption{\textbf{Determination of the anomalous contribution to the elastocaloric effect. }Plot of the anomalous contribution to the elastocaloric effect, $\eta_\textrm{an}$, for strains applied along the $c$-axis (a) and the $b$-axis (b). The inset of (b) shows the background that was subtracted from the measured $\eta$ to obtain $\eta_\textrm{an}$.}
\label{fig:SI-background}
\end{figure}

In Fig. 3 of the main text, we display a color map of the anomalous contribution to the elastocaloric effect, $\eta_\textrm{an}$. In fig.~\ref{fig:SI-background}, we provide a plot of $\eta_\textrm{an}$ as a function of temperature, $T$, and outline the method used to extract $\eta_\textrm{an}$ from the measured elastocaloric data, $\eta$. To this end, we fitted the high-temperature, smoothly-changing $\eta(T)$ vs. $T$ data with a polynomial constrained to pass through zero at $T~=~0$ (see the inset of fig.~\ref{fig:SI-background}). For the $c$-axis data, particularly under high compression, we fitted data above approximately 25~K. For the $b$-axis data, the fit was applied to data above 15~K.

\clearpage
\subsection*{\underline{\small Estimated changes of the triangular-lattice Hubbard model parameters with $b$- and $c$-axis strain}}

Here, we outline our approach to estimate the changes in model parameters for the effective-dimer triangular-lattice Hubbard model under $b$- and $c$-axis strains, $\varepsilon_\textrm{b}$ and $\varepsilon_\textrm{c}$. Within this model, the key parameters are the geometric frustration ratio, $t^\prime/t$, and the effective correlation strength, $U/W$ (with $W$ being the electronic bandwidth). These parameters are generally derived from density-functional theory calculations (e.g., Ref.~\citen{Kandpal09}), which often use experimentally obtained structural data. Currently, no structural data exists for $\kappa$-(ET)$_2$Cu$_2$(CN)$_3$ under strain. However, the evolution of the crystal structure with temperature and corresponding changes in model parameters were reported in Ref.~\citen{Jeschke12}, which may be used to estimate $\frac{t^\prime}{t}(\varepsilon_\textrm{i})$ and $\frac{U}{W} (\varepsilon_\textrm{i})$, where $i = b, c$.

\begin{figure}[h!]
\centering
\includegraphics[width=.8\columnwidth]{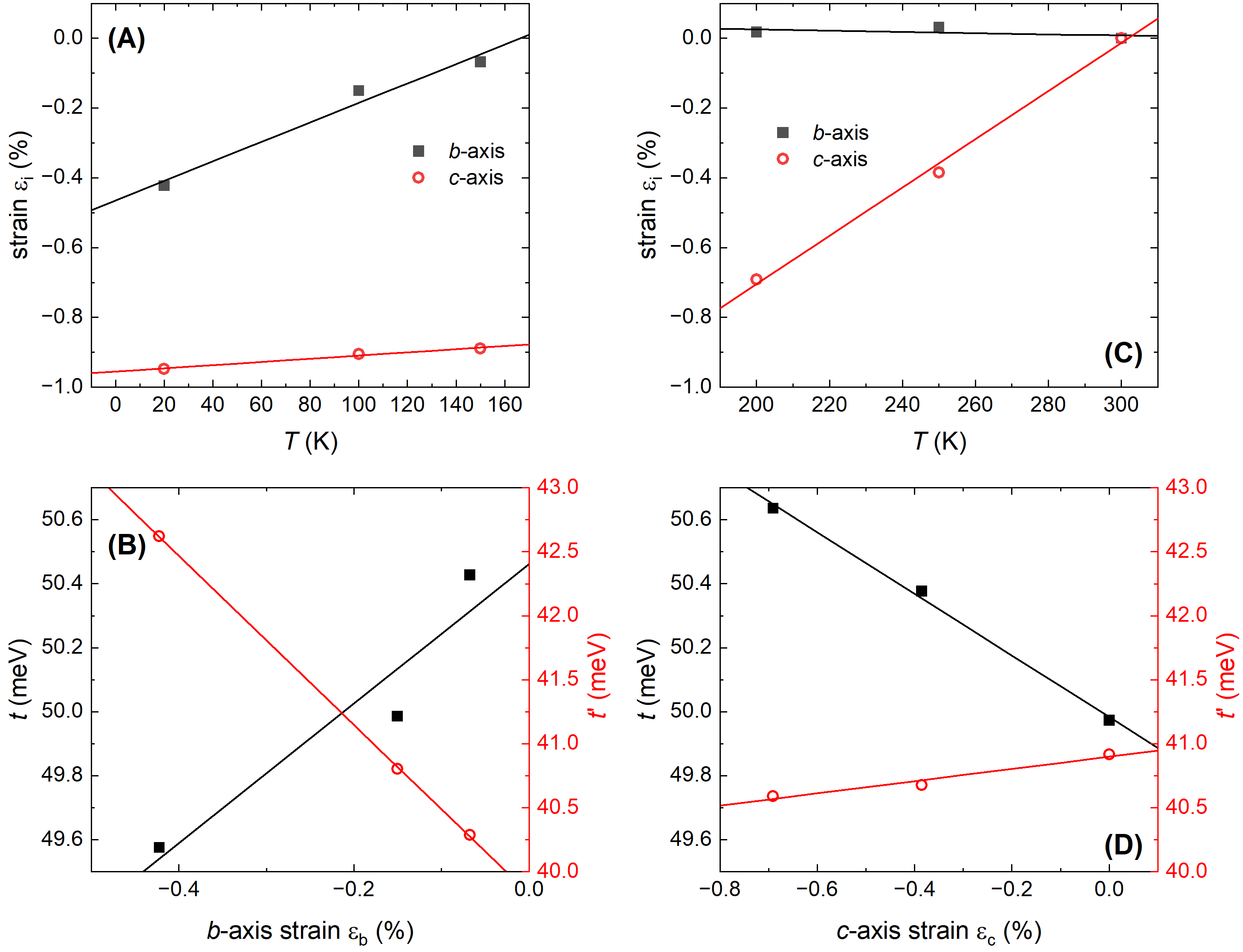} 
\caption{\textbf{Estimation of the geometric frustration ratio, $ t^\prime/t$, under finite strain in $\kappa$-(ET)$_2$Cu$_2$(CN)$_3$, based on density-functional theory calculations from Ref.~\cite{Jeschke12}}. These calculations use the temperature-dependent structure of $\kappa$-(ET)$_2$Cu$_2$(CN)$_3$ at ambient pressure (see text for further details). All strain values are calculated with respect to the room-temperature lattice constants, i.e., $\varepsilon_\textrm{i}~=~(x/x(300~$K$)-1)$ with $x=b,c$ the in-plane lattice parameters. Panels (a) and (b) show the low-temperature evolution of the $ b $- and $ c $-axis lattice strain (a) and the corresponding calculated changes in $ t $ and $ t^\prime $ (b); panels (c) and (d) display the high-temperature evolution of the $b$- and $ c $-axis lattice strain (c) and the calculated variations in $ t $ and $t^\prime$ (d). Solid lines represent linear fits to the data.}
\label{fig:SI-strain-frustration}
\end{figure}

We first focus on $\frac{t^\prime}{t}(\varepsilon_\textrm{i})$. A close inspection of the in-plane lattice constants (or, equivalently, lattice strains, $\varepsilon_\textrm{i}~=~(x/x(300~\textnormal{K})-1)$ with $x=b,c$ the unit-cell lattice parameters) as a function of temperature reveals two distinct regimes (see fig.~\ref{fig:SI-strain-frustration}): at low temperatures ($T \lesssim 150$~K), $\varepsilon_\textrm{b}$ exhibits significant variation, while $\varepsilon_\textrm{c}$ remains relatively stable (see fig.~\ref{fig:SI-strain-frustration}~(a)). Conversely, at high temperatures ($T \gtrsim 200$~K), $\varepsilon_\textrm{b}$ is nearly constant, whereas $\varepsilon_\textrm{c}$ shows substantial changes (see fig.~\ref{fig:SI-strain-frustration}~(c)). Based on this, we use the low-temperature (high-temperature) data to estimate the changes in $t$ and $t^\prime$ with respect to $\varepsilon_\textrm{b}$ ($\varepsilon_\textrm{c}$). This analysis, shown in fig.~\ref{fig:SI-strain-frustration}~(b) and (d), provides:

\begin{eqnarray}
\frac{\textnormal{d}t}{\textnormal{d}\varepsilon_\textrm{b}} &=& 218~\textnormal{meV} \\
\frac{\textnormal{d}t^\prime}{\textnormal{d}\varepsilon_\textrm{b}} &=& -660~\textnormal{meV} \\
\frac{\textnormal{d}t}{\textnormal{d}\varepsilon_\textrm{c}} &=& -96~\textnormal{meV} \\
\frac{\textnormal{d}t^\prime}{\textnormal{d}\varepsilon_\textrm{c}} &=& 48~\textnormal{meV}. 
\end{eqnarray}

\noindent To estimate the change in frustration when uniaxial pressure is applied along a specific axis (e.g., the $c$-axis), we need to account for the fact that this pressure also induces strains along the $b$-axis, so that the $b$-axis strain is a function of the $c$-axis strain. The strain dependence of $t$ and $t^\prime$, when pressure is applied along the $c$-axis, can then be expressed as

\begin{eqnarray}
t (\varepsilon_\textrm{c}) &=& t_0+ \frac{\textnormal{d}t}{\textnormal{d}\varepsilon_\textrm{b}} \varepsilon_\textrm{b}(\varepsilon_\textrm{c}) + \frac{\textnormal{d}t}{\textnormal{d}\varepsilon_\textrm{c}} \varepsilon_\textrm{c} \\
t^\prime (\varepsilon_\textrm{c}) &=& t^\prime_0+ \frac{\textnormal{d}t^\prime}{\textnormal{d}\varepsilon_\textrm{b}} \varepsilon_\textrm{b}(\varepsilon_\textrm{c}) + \frac{\textnormal{d}t^\prime}{\textnormal{d}\varepsilon_\textrm{c}} \varepsilon_\textrm{c},
\end{eqnarray}

\noindent where $t_0$ and $t^\prime_0$ are the values at zero strain, and similarly when stress is applied along the $c$-axis. The magnitude of the strain perpendicular to the primary stress direction is determined by Poisson's ratio $\nu_1 = - \varepsilon_\textrm{b} / \varepsilon_\textrm{c}$ for pressure along the $c$-axis and $\nu_2 = - \varepsilon_\textrm{c} / \varepsilon_\textrm{b}$ for pressure along the $b$-axis. Using Poisson ratios we can rewrite the above equation as 

\begin{eqnarray}
t (\varepsilon_\textrm{c}) &=& t_0+ \left(\frac{\textnormal{d}t}{\textnormal{d}\varepsilon_\textrm{c}} - \nu_1 \frac{\textnormal{d}t}{\textnormal{d}\varepsilon_\textrm{b}} \right) \varepsilon_\textrm{c} \\
t^\prime (\varepsilon_\textrm{c}) &=& t^\prime_0+ \left(\frac{\textnormal{d}t^\prime}{\textnormal{d}\varepsilon_\textrm{c}} - \nu_1 \frac{\textnormal{d}t^\prime}{\textnormal{d}\varepsilon_\textrm{b}}\right) \varepsilon_\textrm{c} \\
t (\varepsilon_\textrm{b}) &=& t_0+ \left(\frac{\textnormal{d}t}{\textnormal{d}\varepsilon_\textrm{b}} - \nu_2 \frac{\textnormal{d}t}{\textnormal{d}\varepsilon_\textrm{c}} \right) \varepsilon_\textrm{b} \\
t^\prime (\varepsilon_\textrm{b}) &=& t^\prime_0+ \left(\frac{\textnormal{d}t^\prime}{\textnormal{d}\varepsilon_\textrm{b}} - \nu_2 \frac{\textnormal{d}t^\prime}{\textnormal{d}\varepsilon_\textrm{c}}\right) \varepsilon_\textrm{b}. \\ 
\end{eqnarray}

\noindent or a typical material, Poisson's ratio is approximately 1/3, which we adopt here as well. Using this value along with the low-temperature values of $ t_0 $ and $ t^\prime_0 $, we obtain the following results for the strain dependence of $t$ and $t^\prime$:

\begin{eqnarray}
t (\varepsilon_\textrm{b}) [\textnormal{meV}] &=& 49.4+ 250 \varepsilon_\textrm{b} \\
t^\prime (\varepsilon_\textrm{b}) [\textnormal{meV}] &=& 42.5 -676 \varepsilon_\textrm{b} \\ 
t (\varepsilon_\textrm{c}) [\textnormal{meV}] &=& 49.4-169 \varepsilon_\textrm{c} \\
t^\prime (\varepsilon_\textrm{c}) [\textnormal{meV}] &=& 42.5+ 268 \varepsilon_\textrm{c}.
\end{eqnarray}

\noindent In the effective-dimer triangular-lattice Hubbard model, correlations are typically quantified by $U/W$, with $U$ the on-site Coulomb repulsion and $W~\propto t,t^\prime$ the electronic bandwidth. $U/W$ can be effectively tuned by hydrostatic pressure (i.e., volumetric strains), since it increases both $t$ and $t^\prime.$ In contrast, anisotropic strains affect $t$ and $t^\prime$ in opposite ways and are therefore not expected to be as effective in tuning $W$. To estimate the changes in correlation strength, we consider the changes of $U/t$ and $U/t^\prime$ with volumetric strains, $\varepsilon_V$ from the data of Ref.\citen{Jeschke12}. For the maximum volumetric strain in our experiments, $\varepsilon_V~\approx~-0.3~\%$ (which is $ 1/3 \cdot \varepsilon_{\textrm{max}}$ due to the Poisson ratio), both $U/t$ and $U/t^\prime$ change by at most 2~\%. Larger changes of the correlation strength are needed to induce the Mott metal-insulator transition~\cite{Pustogow18}. This is consistent with the experimental observation that 1.5~kbar of hydrostatic pressure are needed to induce the Mott metal-insulator transition, i.e., a pressure higher than the maximum uniaxial pressure of 1~kbar applied in the present work. Previous resistance measurements under uniaxial pressure on $\kappa$-(ET)$_2$Cu$_2$(CN)$_3$~\cite{Shimizu11} also confirm that higher uniaxial than hydrostatic pressures are needed to induce the Mott metal-insulator transition.

\clearpage
\subsection*{\underline{\small Modelling of ECE and phase diagrams}}

\noindent \textit{Landau free energy model - } In the following, we introduce in detail the Landau free energy model, which we refer to in the main text. We begin by shortly reviewing the results for the phase diagram of two competing order parameters, $\Phi_1$ and $\Phi_2$. In this case, the Landau free energy reads as

\begin{equation}
    \mathcal{F} = a_1 \Phi_1^2 + a_2 \Phi_1^4 + b_1 \Phi_2^2 + b_2 \Phi_2^4+ \lambda_1 \Phi_1^2\Phi_2^2 + \lambda_2 \Phi_1^2\Phi_2^4,
\end{equation}

\noindent with $a_1~\propto~(T-T_1^0)$ and $b_1~\propto~(T-T_2^0)$ that depend on the distance to the non-interacting transition temperatures $T_1^0$ and $T_2^0$ and $a_2, b_2 $ being the prefactors of the quartic terms required for stability. The interaction strength $\lambda_1$ can either be positive (repulsive interaction) or negative (attractive interaction). For a large attractive interaction, the term $\lambda_2 \Phi_1^2\Phi_2^4$ ($\lambda_2~>~0$) is needed for stability (note that we omit the term $\Phi_1^4\Phi_2^2$ for simplicity without loss of generality). In order to model a phase diagram as a function of strain, we consider that $T_1^0$ and $T_2^0$ are strain-dependent.

\begin{figure}[h!]
\centering
\includegraphics[width=.8\columnwidth]{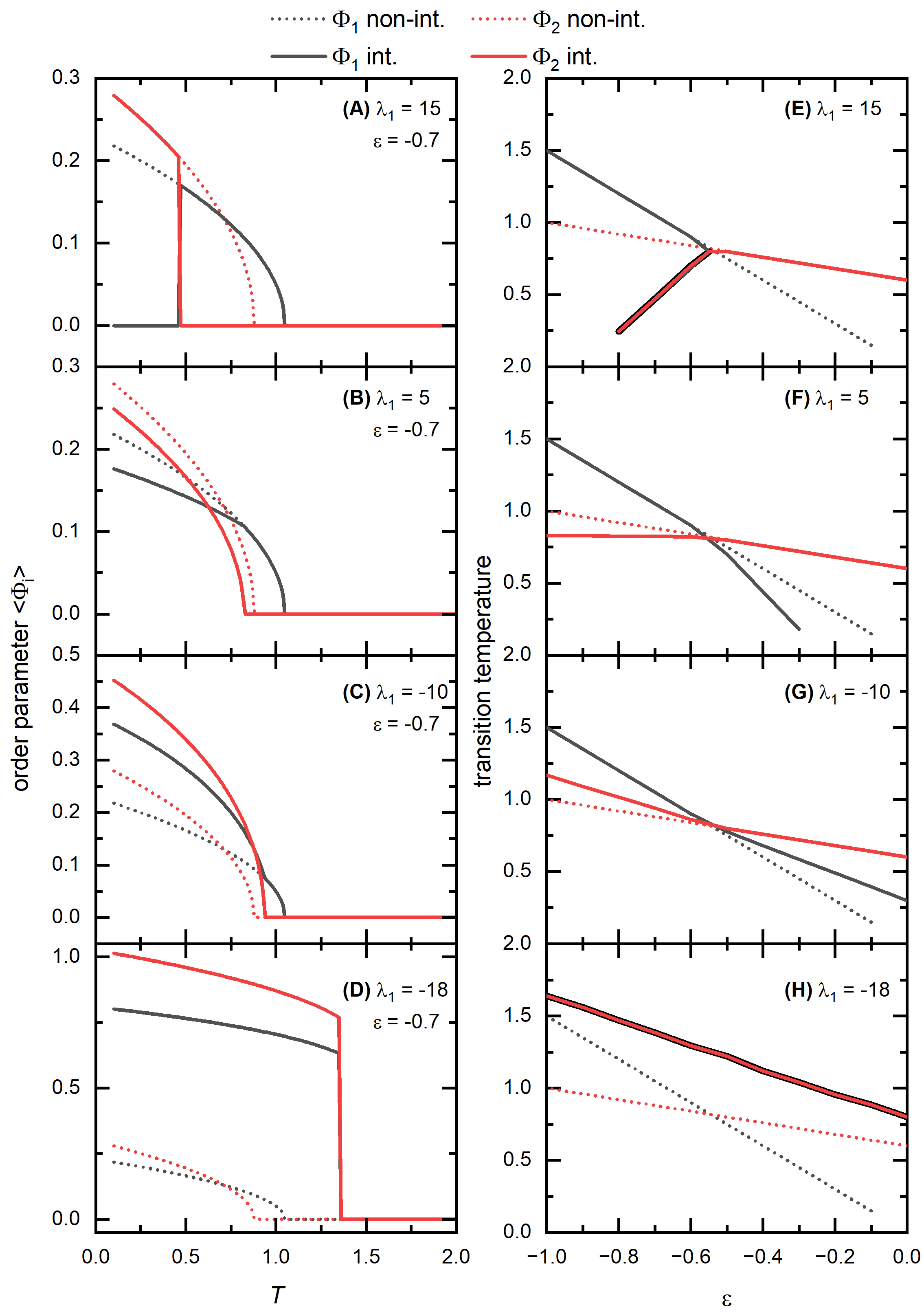} 
\caption{\textbf{Results of a Landau free energy modelling for two order parameters ($\Phi_1$ and $\Phi_2$) with varying interaction strength, $\lambda_1$}. For these simulations, we vary $\lambda_1$ between 15 and $-$18, corresponding to situations with strong repulsive to strong attractive interactions. The left column (\textbf{A}-\textbf{D}) depicts the temperature dependence of $\Phi_1$ and $\Phi_2$ at different $\lambda_1$ (solid lines). The right column (\textbf{E}-\textbf{H}) shows the calculated phase diagrams as a function of strain, $\varepsilon$, for the same $\lambda_1$ (solid lines). In each panel, the results for the non-interacting case ($\lambda_1~=\lambda_2~=~$0) are included as dotted lines for comparison. For the calculations, we used $a_1~=~(T+0.4(\varepsilon-1.5))$, $a_2~=~5$, $b_1~=~(T+1.5\varepsilon)$, $b_2~=~10$ and $\lambda_2~=~5$. By definition, temperature and strain are unitless parameters in these calculations.}
\label{fig:SI-2OPmodelling}
\end{figure}

In fig.~\ref{fig:SI-2OPmodelling}, we illustrate the temperature dependence of the two order parameters (for a strain at which $T_1^0~>~T_2^0$) and the resulting temperature-strain phase diagrams for varying degrees of interaction, $\lambda_1$. For strong competing interaction (see fig.~\ref{fig:SI-2OPmodelling}~(a) and (e)), there is no coexistence region, where both $\Phi_1$ and $\Phi_2$ orders coexist. The temperature-induced $\Phi_1$-$\Phi_2$ transition is first order. The slope of this first-order line is markedly different from the slopes of the non-interacting $\Phi_1$ and $\Phi_2$ phase transition lines. In case of moderate competition (see fig.~\ref{fig:SI-2OPmodelling}~(b) and (f)), the onset of $\Phi_2$ does not entirely eliminate $\Phi_1$, but weakens it. The corresponding phase diagram has four transition lines that merge in a polycritical point, at which the slopes of the phase transition lines change. The region of coexisting $\Phi_1$ and $\Phi_2$ order parameters is smaller than for the non-interacting case. For moderate attractive interactions (see fig.~\ref{fig:SI-2OPmodelling}~(c) and (g)), the presence of $\Phi_2$ strengthens $\Phi_1$ even further. The phase diagram still exhibits a polycritical point with four lines merging, but now the region of coexisting $\Phi_1$/$\Phi_2$ order is enhanced compared to the non-interacting case. Finally, for strong attractive interactions (see fig.~\ref{fig:SI-2OPmodelling}~(d) and (h)), the $\Phi_1$ and $\Phi_2$ transition merge into a single first-order transition.

The two-order parameter model can be used to discuss the implications for a VBS+CO state for $\kappa$-(ET)$_2$Cu$_2$(CN)$_3$ at zero strain.  As discussed in the main text, the analysis of the symmetries of the VBS and the CO state establishes both order parameters as distinct.  Since the transition temperatures for both orders are identical, CO and VBS must couple strongly attractively and the coupled VBS+CO transition must be first order.

If the $T^\star$ anomaly would correspond to only VBS order, then the interplay of the VBS order with the new order should give rise to a topology of the phase diagrams as shown in fig.~\ref{fig:SI-2OPmodelling}~(e), (f), (g) or (h). However, our experimental phase diagram with (i) only three phase transition lines that meet at a polycritical point and (ii) a basically unchanged slope of the $T^\star$ transition line at the polycritical point does not match any of the model phase diagrams.

\begin{figure}[h!]
\centering
\includegraphics[width=\columnwidth]{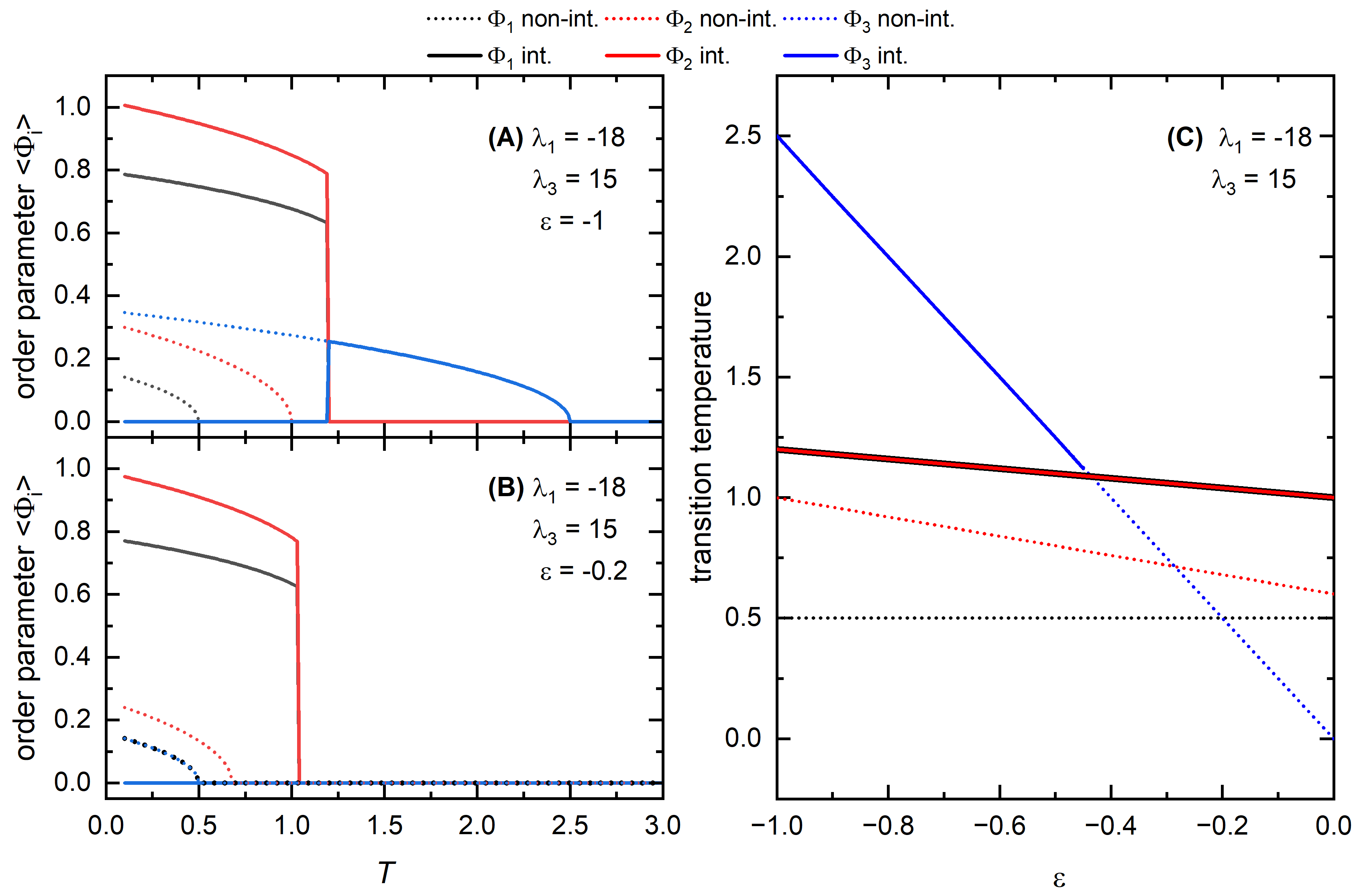} 
\caption{\textbf{Results of a Landau free energy modelling for three order parameters ($\Phi_1$, $\Phi_2$ and $\Phi_3$), with $\Phi_1$ and $\Phi_2$ interacting attractively ($\lambda_1~=~-18$) and $\Phi_1$ and $\Phi_2$ repelling each other ($\lambda_3~=~15$).} The left column (\textbf{A}, \textbf{B}) depicts the temperature dependence of the three order parameters at two different strains. The right column (\textbf{C}) shows the calculated phase diagram as a function of strain, $\varepsilon$, for this set of parameters. In each panel, the results for the non-interacting case ($\lambda_1~=\lambda_2~=~\lambda_3~=~$0) are included as dotted lines for comparison. The calculation used $a_1~=~(T+0.4(\varepsilon-1.5))$, $a_2~=~5$, $b_1~=~T-0.5$, $b_2~=~10$, $c_1~=~(T+2.5\varepsilon)$, $c_2~=~10$ and $\lambda_2~=~5$. By definition, temperature and strain are unitless parameters in these calculations.}
\label{fig:SI-3OPmodelling}
\end{figure}

Thus, we extend our model discuss the case of three order parameters, where a coupled VBS+CO order at zero strain interacts with a third order parameter (corresponding to the new phase) at high compression. In this case, the free energy reads as

\begin{equation}
    \mathcal{F} = a_1 \Phi_1^2 + a_2 \Phi_1^4 + b_1 \Phi_2^2 + b_2 \Phi_2^4 + c_1 \Phi_3^2+ c_2 \Phi_3^4 + \lambda_1 \Phi_1^2\Phi_2^2 + \lambda_2 \Phi_1^2\Phi_2^4 + \lambda_3 \Phi_2^2 \Phi_3^2,
\end{equation}

\noindent with $c_1~\propto(T-T_3^0)$ the interaction parameter $\lambda_3$ between $\Phi_2$ and $\Phi_3$. For our calculations, we chose $\lambda_3$ to be positive. This choice is motivated by the physical scenario in which antiferromagnetic order emerges at low frustration ($\Phi_3$), which competes with the VBS order (for simplicity and without loss of generality, we omit other higher-order terms $\Phi_i^4\Phi_j^2$ as well as an interaction between $\Phi_1$ and $\Phi_3$). Following the discussion above, the VBS order and the CO ($\Phi_1$) are taken to strongly attract one another. In figs.~\ref{fig:SI-3OPmodelling}~(a) and (b), we compare the temperature dependence of the order parameter for $T_3^0~>T_2^0,T_1^0$ (case a) and $T_2^0~>T_3^0,T_1^0$ (case b). In case (a), upon cooling, $\Phi_3$ condensed first through a second-order phase transition. Upon further cooling, the system undergoes a strong first-order transition to an $\Phi_1$+$\Phi_2$ state. In case (b), there is also a first-order $\Phi_1$+$\Phi_2$ transition upon cooling, but $\Phi_3$ never undergoes a transition. Correspondingly, the phase diagram (fig.~\ref{fig:SI-3OPmodelling}~(c)) shows a polycritical point, at which the second-order $\Phi_3$ line and the first-order $\Phi_1$+$\Phi_2$ transition line merge. We note that the slope of the $\Phi_1$+$\Phi_2$ line remains essentially unchanged across the tricritical point. \\

\noindent \textit{Elastocaloric effect at a first-order phase transition - }In the following, we discuss the characteristic signatures of a first-order transition in elastocaloric effect (ECE) measurements. At an ideal first-order phase transition, the entropy $ S $ exhibits a discontinuity as a function of both strain ($ \varepsilon $) and temperature ($ T $). Consequently, the derivatives $ \partial S / \partial \varepsilon $ and $ \partial S / \partial T $ diverge, rendering the ECE, defined as $ \eta_\textrm{i} = \left(\frac{\Delta T}{\Delta \varepsilon_\textrm{i}}\right)_S = -\frac{\left(\frac{\partial S}{\partial \varepsilon_\textrm{i}} \right)_T}{\left( \frac{\partial S}{\partial T}\right)_{\varepsilon_\textrm{i}}} $, ill-defined. 

\begin{figure}[h!]
\centering
\includegraphics[width=.8\columnwidth]{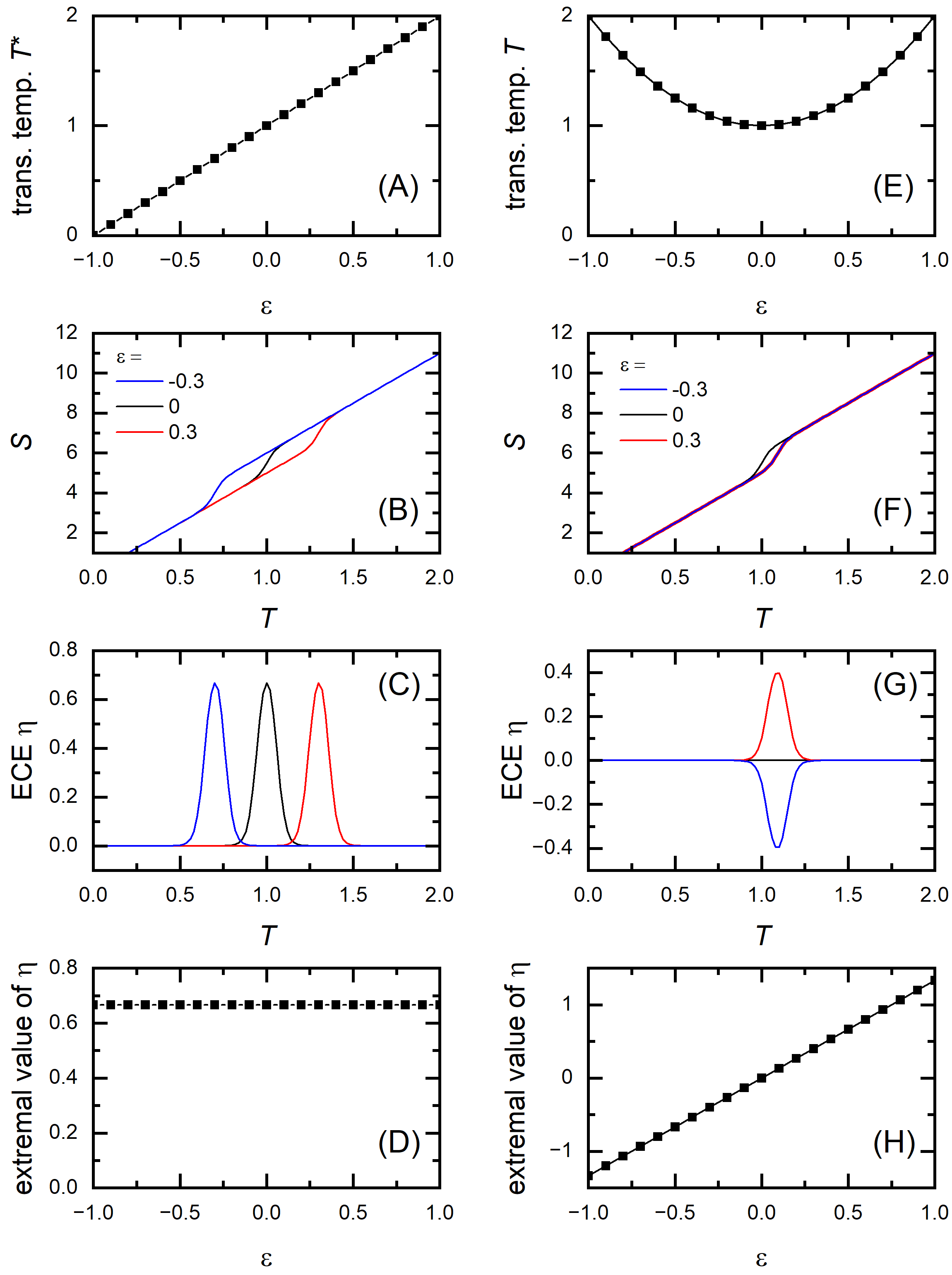} 
\caption{\textbf{Calculations of the elastocaloric effect (ECE) across a broadened first-order transition.} (a-d) The ECE for a broadened first-order transition whose transition temperature, $T^\star$, changes linearly with strain (a). The model entropy, $S$, (b) consists of a broadened entropy jump, $\Delta S$, and an in $T$ linear background. The corresponding ECE (c) shows a large, symmetric peak. (d) The maximum value of the ECE is independent of strain, just as d$T^\star$/d$\varepsilon$. (e-h) Same calculations the case when the transition temperature is minimal at zero strain. The anomaly in the temperature-dependent ECE has a different sign for $\varepsilon~<~0$ and $\varepsilon~>~0$, and the extremal value of ECE follows a linear behavior, just as d$T^\star$/d$\varepsilon$.}
\label{fig:SI-firstordermodelling}
\end{figure}

However, in real systems, the entropy jump at $ S(T, \varepsilon) $ is not infinitely sharp but broadened, ensuring that $ \partial S / \partial \varepsilon $ and $ \partial S / \partial T $ remain finite. Additionally, there is a background contribution to the entropy, such as from phonons, whose dependence on $ T $ and $ \varepsilon $ differs from the degrees of freedom involved in the phase transition. As a result, the ECE can still be evaluated for a broadened first-order transition, as illustrated in fig.~\ref{fig:SI-firstordermodelling}. 

For the calculations shown, a background entropy linear in $T$, independent of strain, was assumed. Figure~\ref{fig:SI-firstordermodelling}(a-d) depict the calculated ECE for a broadened first-order phase transition where the ordering temperature, $T^\star$, varies linearly with strain. The ECE displays a symmetric peak at the transition temperature. The maximum value of the ECE at each strain is independent of strain: since the ECE is proportional to the strain derivative of the entropy, it is also related to d$T^\star$/d$\varepsilon$, which is constant in the present case.

In fig.~\ref{fig:SI-firstordermodelling}(e-h), we explore a scenario where the transition temperature $ T^\star(\varepsilon) $ exhibits a minimum at zero strain.  Here, the ECE anomaly is large and positive at $\varepsilon~=~1$, decreases as $\varepsilon$ is reduced, crosses zero at $ \varepsilon = 0 $, and becomes strongly negative at $\varepsilon~=~-1$. This behavior is linked to the change in sign of $ \partial S / \partial \varepsilon $ at $ \varepsilon~=~0$. In this case the extremal value of the ECE linearly depends on $\varepsilon$, changing sign at the point of maximal entropy. This scenario bears a strong resemblance to our measured $\eta_\textrm{an}$ for strain along the b-axis, shown in the main text.



\end{document}